\documentclass[usenatbib]{mnras}
\usepackage{graphicx, longtable}

\newcommand\msun{{\rm M_{\odot}}}

\usepackage[hyphenbreaks]{breakurl}

\def\go{
\mathrel{\raise.3ex\hbox{$>$}\mkern-14mu\lower0.6ex\hbox{$\sim$}}
}
\def\lo{
\mathrel{\raise.3ex\hbox{$<$}\mkern-14mu\lower0.6ex\hbox{$\sim$}}
}
\title[V3890 Sgr]
{The 2019 eruption of recurrent nova V3890 Sgr: observations by {\em Swift}, {\em NICER} and SMARTS}
\author[K.L. Page et al.]{K.L. Page$^{1}$\thanks{E-mail: klp5@leicester.ac.uk (KLP)}, N.P.M. Kuin$^{2}$, A.P. Beardmore$^{1}$, F.M. Walter$^{3}$, J.P. Osborne$^{1}$, \newauthor C.B. Markwardt$^{4}$, J.-U. Ness$^{5}$, M. Orio$^{6,7}$ and K.V. Sokolovsky$^{8,9}$ \\
$^{1}$ X-Ray and Observational Astronomy Group, School of Physics \&
  Astronomy, University of Leicester, LE1 7RH, UK\\
  $^{2}$ Mullard Space Science Laboratory, University College London, Holmbury St. Mary, Dorking, Surrey RH5 6NT, UK\\
$^{3}$  Department of Physics and Astronomy, Stony Brook University, Stony Brook, NY 11794-3800, USA\\  
$^{4}$  Astrophysics Science Division, NASA Goddard Space Flight Center, Greenbelt, MD 20771, USA\\
$^{5}$ European Space Astronomy Centre, E-28692 Villanueva de la Ca{\~ n}ada, Madrid, Spain\\
$^{6}$ INAF-Osservatorio di Padova, Vicolo Osservatorio 5, 35122 Padova, Italy\\
$^{7}$ Department of Astronomy, University of Wisconsin, 475 N. Charter St., Madison WI 53706, USA\\
  $^{8}$ Center for Data Intensive and Time Domain Astronomy, Department of Physics and Astronomy, Michigan State University, \\East Lansing, MI 48824, USA\\
$^{9}$ Sternberg Astronomical Institute, Moscow State University, Universitetskii pr. 13, 119992 Moscow, Russia\\
}

\date{Accepted XXX. Received YYY; in original form ZZZ}

\pubyear{2020}

\begin{document}
\label{firstpage}
\pagerange{\pageref{firstpage}--\pageref{lastpage}}

\maketitle

\begin{abstract}

  V3890 Sgr is a recurrent nova which has been seen in outburst three times so far, with the most recent eruption occurring on 2019 August 27 UT. This latest outburst was followed in detail by the {\em Neil Gehrels Swift Observatory}, from less than a day after the eruption until the nova entered the Sun observing constraint, with a small number of additional observations after the constraint ended. The X-ray light-curve shows initial hard shock emission, followed by an early start of the super-soft source phase around day 8.5, with the soft emission ceasing by day 26. Together with the peak blackbody temperature of the super-soft spectrum being $\sim$~100~eV, these timings suggest the white dwarf mass to be high, $\sim$~1.3$\msun$. The UV photometric light-curve decays monotonically, with the decay rate changing a number of times, approximately simultaneously with variations in the X-ray emission. The UV grism spectra show both line and continuum emission, with emission lines of N, C, Mg and O being notable. These UV spectra are best dereddened using an SMC extinction law. Optical spectra from SMARTS show evidence of interaction between the nova ejecta and wind from the donor star, as well as the extended atmosphere of the red giant being flash-ionized by the super-soft X-ray photons. Data from {\em NICER} reveal a transient 83~s quasi-periodic oscillation, with a modulation amplitude of 5~per~cent, adding to the sample of novae which show such short variabilities during their super-soft phase.

\end{abstract}


\section{Introduction}
\label{intro}

Novae are caused by thermonuclear explosions within an interacting binary system. Those termed classical novae (CNe) have been observed in outburst only once, while a small minority, the recurrent novae (RNe), have been detected in outburst more often, likely due to their higher white dwarf (WD) masses and accretion rates \citep{brad10}. In CNe, which typically have orbital periods of hours, the companion star is usually on the main sequence, whereas the RNe often have much larger orbits, with the secondaries evolved to sub- or red giants \citep{darn12}.

Continued mass transfer eventually ignites nuclear burning at the base of the accreted envelope; this then causes the pressure to reach a  sufficient level to trigger a thermonuclear runaway \cite[TNR; see][for review articles]{bode08,woudt14}. Immediately following this explosion, the ejected material obscures the WD surface from view. As the ejecta expand however, they become optically thin, typically allowing the surface nuclear burning to become visible. The ejecta photosphere recedes back to that of the WD which may still be heated by residual nuclear burning and, so, emitting soft X-rays. This stage is known as the Super-Soft Source (SSS) state \citep{kraut08}, and has been well observed in many novae by the {\em Neil Gehrels Swift Observatory} \citep[henceforth, {\em Swift}; ][]{geh04}; see \citet*{julo15, page20} for recent summaries of these observations.

There are ten confirmed Galactic RNe, of which V3890~Sgr is one, each with recurrence timescales of $\sim$~10--100~yr, although the upper bound for these recurrence times is obviously a selection effect dependent on historical records\footnote{RNe outside the Milky Way have been found with recurrence times as short as 1~yr \cite[e.g.][]{darn14}.}; \citet{brad10}  provides a review of the recurrent systems in the Milky Way. 

V3890~Sgr was previously known to have erupted in 1962 and 1990 \citep{1962, 1990, brad10}. On 2019 August 27.87 UT, the system was noted to be in outburst for a third time \citep{per19}, confirming the recurrence interval to be 28--29 yr. An observation on 2019 August 27.05 by ASAS-SN (All-Sky Automated Survey for SuperNovae) showed the source still to be in quiescence, while an observation on August 27.75 was saturated \citep{strader19}. A first measured magnitude of V~=~7.17 was obtained on August 28.1188 \citep{sok19}. For the plots presented in this paper, 2019 August 27.75 will be assumed to be T0, but there remains an uncertainty of 0.7 days for the exact outburst time, between the final non-detection and the saturated exposure.

\citet{rob06} performed a search for additional outbursts of V3890~Sgr, but none was found. They note that the object is moderately fast in decline, and behind the Sun each December, so there is scope for eruptions to be missed entirely.

V3890~Sgr is one of the subsection of the RNe known as `symbiotic-like recurrents', where the mass-losing star is a red giant (RG), making it similar to RS~Oph and V407~Cyg \citep{brad10}. The system contains an M5 {\sc iii} RG secondary star \citep{harr93} which is pulsating with a $\sim$~104~day periodicity \citep{brad10,mroz14}. \citet{brad09} finds an orbital period of 519.7~$\pm$~0.3~day, based on photometry spanning more than a century.

For this 2019 eruption, the nova was also identified by the names AT~2019qaq and Gaia19ebf\footnote{See \url{https://wis-tns.weizmann.ac.il/object/2019qaq}}. The outburst was detected across the electromagnetic spectrum, from $\gamma$-rays (\citealt{buson19}), through X-rays \citep{sok19,orio19,page19a, page19b,beardmore19a,beardmore19b,ness19b,singh19a,singh19b, orio20} and UV \citep{kuin19} to optical \citep{strader19,mun19,mun19b}, infrared \citep{evans19,wood19,wood20} and radio \citep{nya19,pol19}. Here we present a detailed analysis of the {\em Swift}, {\em NICER} \citep[{\em Neutron Star Interior Composition Explorer} ; ][]{nicer, nicer16} and a portion of the SMARTS\footnote{\url{http://www.astro.yale.edu/smarts/}} (Small and Moderate Aperture Research Telescope System) data of the 2019 eruption of V3890~Sgr.

The distance to V3890~Sgr has been taken to be 4.5~kpc, which is the lower limit from \citet{mun19} based on their calculated reddening and the dust map from \citet{green19}, with similar results from using the \citet{lal14} map. \citet{brad09} lists an average estimated distance of 6~$\pm$~1~kpc (measurements ranging from 4.7 to 7.6~kpc), which would be consistent with this lower limit.
\citet{brad18} points out that the parallaxes from {\em Gaia} DR2 are not yet reliable for systems with long-period binary orbits, such as V3890~Sgr. 

 In this paper errors are given at the 90~per~cent confidence level,
 unless otherwise stated. Spectra were binned such that they have a
 minimum of 1 count~bin$^{-1}$ to facilitate Cash statistic
 \citep{cash79} fitting within {\sc xspec} \citep[strictly speaking, a modified version for the case where there is a background file\footnote{\url{https://heasarc.gsfc.nasa.gov/docs/xanadu/xspec/manual/XSappendixStatistics.html}};][]{arn96}, and the
 abundances from \citet{wilms00}, together with the photoelectric
 absorption cross-sections from \citet{vern96}, have been assumed when
 using the T{\" u}bingen-Boulder Interstellar Medium ({\sc xspec/tbabs}) absorption model.

\section{Observations}
\label{obs}

\subsection{\em Swift}
{\em Swift} observations of V3890~Sgr began less than 0.7 days after the assumed eruption time of 2019 August 27.75, and a mere 0.5 days after the announcement of the discovery, collecting data using both the 
X-ray Telescope \citep[XRT; ][]{bur05} and UV/Optical Telescope \citep[UVOT; ][]{rom05}. A bright, hard X-ray source was immediately detected, with a corresponding bright UV source \citep{sok19}. Multiple snapshots of data\footnote{A continuous pointing by {\em Swift} is known as a snapshot.}  were collected each day between 2019 August 28 and September 05 (days 0.7--9 after the nova outburst), when the SSS emission was first detected \citep{page19a}. At this point, V3890~Sgr became too close to the Moon for {\em Swift} to observe, leading to a five day gap in observations. Data collection recommenced on September 10 (day 13), with several kiloseconds of data collected every day until September 26 (day 30), aimed at constraining any short-term variability. The exposure was then cut to $\sim$~1~ks per day until the end of September (day 34), when the cadence of observations was decreased to approximately every three days. Three additional weekly observations were collected between October 28 and November 11, taking the observations up to 75 days post-outburst, at which time the nova became too close to the Sun in the sky for {\em Swift} to observe for the following three months. Once V3890~Sgr re-emerged from the solar observing constraint, five further weekly observations were obtained, between 2020 February 17 and March 14 (days 173--199), with a final 1.7~ks of data collected between April 29 and May 04 (days 245--250).

Most UVOT observations were performed using the $uvm2$ filter (central wavelength of 2246 \AA). However, a number of UV-grism spectra were also taken between September 02 and 16 (days 5--20; together with $uvw2$ exposures, which is the default for the standard UVOT UV grism mode), with a V-grism observation on October 02 (day 35). In addition, the first four snapshots used the $uvw1$ and $u$ filters. The roll angle was varied between snapshots, and an offset was implemented for the grism observations from day 16 onwards. 
The earliest observations on August 28, 29 and the start of 30 found a source which was too bright for standard UVOT photometry; however, the method of using the read-out streaks \citep{mat13} could be utilised to obtain magnitudes at these times. These measurements are given in Table~\ref{uvot}. In addition, for one of the observations (ObsID 00045788015) the source fell on one of the known areas of the UVOT detector where the throughput is lower\footnote{\url{https://swift.gsfc.nasa.gov/analysis/uvot\_digest/sss\_check.html}}; this dataset was therefore excluded from the UVOT analysis. The full set of UVOT photometry is listed in Table~\ref{photom}. All magnitudes are given in terms of the Vega system.

\begin{table}

\caption{Early UVOT magnitudes derived from the read-out streak. Times are the mid-point of the observation in days since eruption. The error given is statistical only; there is an additional systematic error of $\sim$~0.10 in each case.}

\begin{center}
\begin{tabular}{lcc}
\hline
Day & Filter & magnitude\\
\hline
0.66 & $uvw1$ & 8.59~$\pm$~0.02\\
0.72 & $uvw1$ & 8.62~$\pm$~0.02\\
1.33 & $u$ & 8.84~$\pm$~0.04\\
2.20 & $u$ & 9.19~$\pm$~0.04\\ 
\hline
\end{tabular}

\label{uvot}
\end{center}
\end{table}

The {\em Swift} data were processed and analysed using {\sc heasoft} version 6.26.1, together with the most recent calibration files available in the latter half of 2019. The UVOT grism images were processed with the {\sc uvotpy} python software version 2.3.2 and calibration \citep{paul14, paul15}.
In addition, an updated pre-release gain file for the XRT-Windowed Timing (WT) data was used\footnote{We note that this gain file, created using calibration data from the second half of 2019, significantly improved the agreement between the {\em Swift} and {\em NICER} results. The file is currently being prepared for release.}. There is a known calibration issue for the WT data, referred to as `trailing charge'\footnote{\url{https://www.swift.ac.uk/analysis/xrt/digest_cal.php\#trail}}, which can lead to spurious low-energy bumps in the spectrum. To circumvent this problem, a low-energy cut-off of 0.5~keV (rather than the usual 0.3~keV) was used when generating the light-curve and hardness ratio, and when modelling the pre-SSS phase spectra. By the time the SSS had faded away, the data were being collected in Photon Counting (PC)  mode, where trailing charge is not a problem. In addition, to help further minimise the trailing charge and pile-up, only grade 0 (single pixel) events were used for both XRT modes.

PC data were assumed to be piled-up above $\sim$~0.3~count~s$^{-1}$, and corrected appropriately using annular extraction regions\footnote{Pile-up becomes evident at lower count rates in super-soft sources, making it advisable to exclude more of the PSF than generally suggested for the harder sources -- e.g. \citet{romano06}.}. The source never reached a high enough count rate for the WT data to suffer from pile-up.

\begin{figure}
\begin{center}
\includegraphics[clip, width=8.5cm]{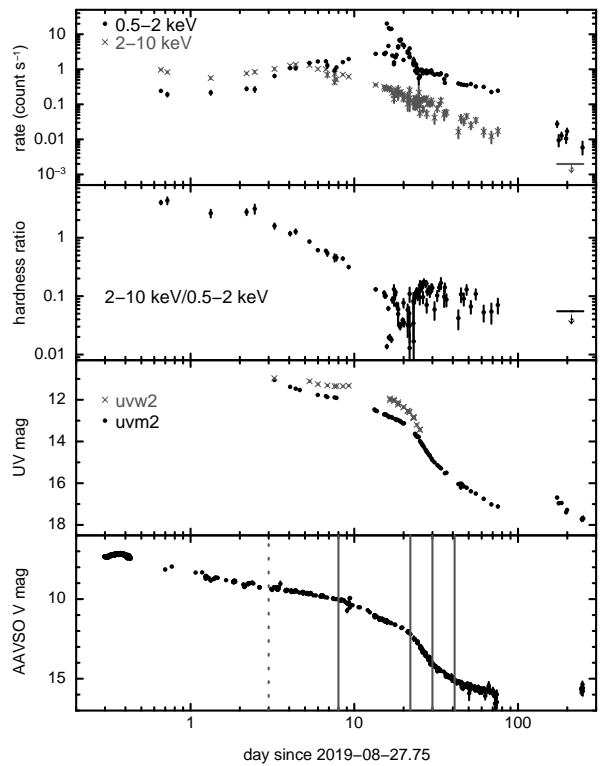}
\caption{From top to bottom: {\em Swift} X-ray soft and hard band light-curves, hardness ratio and UV light-curves, and AAVSO $V$-band light-curves. The vertical grey lines in the bottom panel mark the break times in the decay, as discussed in Section~\ref{breaks}; the dotted line shows the time at which the parametrization of the UV/optical decay begins. Note that the final bins in the hard light-curve and hardness ratio are upper limits.}
\label{lc}
\end{center}
\end{figure}

\begin{figure}
\begin{center}
\includegraphics[clip, angle=-90, width=8.5cm]{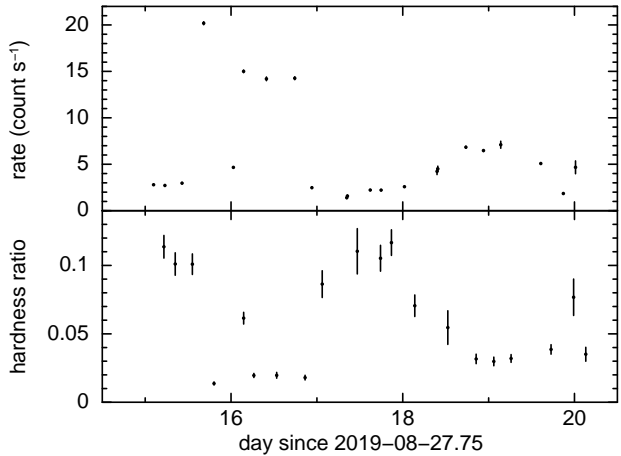}
\caption{A zoom-in of the soft X-ray light-curve and hardness ratio (2-10 keV/0.5-2 keV) from Fig.~\ref{lc}, but plotted in linear space to emphasise the variability. There is a clear anti-correlation between count rate and hardness.}
\label{zoom}
\end{center}
\end{figure}

The {\em Swift} light-curves and X-ray hardness ratio\footnote{Created with one bin per snapshot until after the observing constraint, using the online XRT product generator: \url{https://www.swift.ac.uk/user\_objects/} \citep{evans07,evans09}.}, together with optical data obtained from AAVSO (American Association of Variable Star Observers)\footnote{All available V-band data were included; there was no filtering on uncertainties, and no observers were excluded.} are shown in Fig.~\ref{lc}; Fig.~\ref{zoom} highlights the variability and anti-correlation between the X-ray count rate and corresponding hardness ratio during the early super-soft interval (see Section~\ref{var}). The hardness ratio bands were set as 2--10~keV and 0.5--2~keV, with the cut at 2~keV ensuring all SSS counts were in the lower-energy band. 

For the observations obtained after the source emerged from the solar observing constraint, almost all the photons were below 2~keV; the upper limits shown for the 2--10~keV band and the hardness ratio in Fig.~\ref{lc} combine all the data from this interval.

While the X-ray data show an initial brightening, before starting to fade after around 16 days, the UV and optical light-curves decay monotonically, as frequently seen in {\em Swift}-monitored novae \citep[e.g., ][]{page20}. This is discussed in Section~\ref{disc}.

\subsection{\em NICER}
\label{nicer}

Following the detection of the SSS emission by {\em Swift}, a real-time target of opportunity observation of V3890~Sgr was performed by {\em NICER} on 2019 September 06 \citep[day 9.9 after eruption;][]{beardmore19a}. Following this, a further 13 datasets were collected between September 10 and 27 (days 13--30), with raw exposure times ranging from 0.3~ks to 3.5~ks. However, for a number of these, including the initial dataset, V3890~Sgr was observed as {\em NICER} passed through the South Atlantic Anomaly (SAA) and/or the observations were affected by intervals of high background, meaning that little or no time remained in the default cleaned event-lists.
Relaxing the filtering by removing the {\em nicersaafilt}/{\em saafilt} and {\em overonly\_expr} parameters\footnote{\url{https://heasarc.gsfc.nasa.gov/docs/nicer/data_analysis/nicer_analysis\_guide.html}} recovered much of the exposure from the initial observation on September 06, though had less of an effect on the observations of September 10, 17 and 26. While we did not choose to use these `reclaimed' data for general spectral fitting, due to the possibility of calibration issues, they were useful for the timing analysis given in Section~\ref{var}.

 Appropriate background spectra were extracted following the {\em NICER} background estimator tools\footnote{\url{https://heasarc.gsfc.nasa.gov/docs/nicer/tools/nicer_bkg_est_tools.html}}, and {\sc heasoft} version 6.26.1 was used, together with the most up-to-date calibration files available (release date 2020-02-02, including a gain file update).

\subsection{\em SMARTS}

SMARTS is comprised of a number of telescopes located on Cerro Tololo, Chile. Following the 2019 eruption of V3890~Sgr, 32 spectra were obtained between 2019 August 28 and November 04 (days 1--70), with nightly observations (weather permitting) until September 26, after which the cadence was reduced. All spectra were obtained with the Chiron fibre-fed echelle specrograph \citep{chiron} mounted on the CTIO\footnote{Cerro Tololo Inter-American Observatory} 1.5m telescope, with integration times of 10--60~min, depending on the brightness of the target. The spectra reported here were taken in `fiber mode', with 4$\times$4 on-chip binning yielding a resolution $\lambda/\Delta\lambda \approx$~27,800.

The data were reduced using a pipeline coded in IDL\footnote{\url{http://www.astro.sunysb.edu/fwalter/SMARTS/CHIRON/ch_reduce.pdf}}. The images were flat-fielded. Cosmic rays were removed using the L.A.Cosmic algorithm 
\citep{vanDokkum01}. The 74 echelle orders were extracted, and instrumental background was subtracted. As Chiron is fibre-fed, there is no simple method to subtract the sky. In any event, for bright targets, such as V3890~Sgr in outburst, night sky emission is generally negligible apart from narrow [O\,{\sc i}] and Na {\sc d} lines.

Wavelength calibration was accomplished using ThAr calibration lamp exposures
at the start and end of the night, and occasionally throughout the night.
In our experience, Chiron in `fiber mode' is stable to better than 250 m~s$^{-1}$ over the course of
many nights.
The instrumental response was removed from the individual orders by dividing
by the spectrum of a flux-standard star, $\mu$~Col. This provides 
flux-calibrated orders with a systemic uncertainty due to sky conditions. We 
later account for this by computing broadband flux offsets by calibrating against contemporaneous optical photometry.
The individual orders are spliced together, resulting in a
calibrated spectrum from 4080--8900~\AA. Finally, contemporaneous $BVRI$
photometry from AAVSO was used to scale the spectrum to approximately true fluxes.

The SMARTS observing log is provided in Table~\ref{smartslog}.

\section{X-ray analysis}

\subsection{Variability}
\label{var}

{\em Swift} observations of RS Oph in 2006 first identified a curious phenomenon, whereby the early rise to peak SSS emission was chaotic, with high-amplitude flux variability being seen -- variations in the count rate of an order of magnitude or more in $\sim$~12 hours \citep{julo11}. Detailed monitoring of subsequent novae by {\em Swift} has shown that RS~Oph was not a unique case, with a number of other sources also showing similar variability; \citet{page20} summarise the previous {\em Swift} results. While not as long-lived or spectacular as RS~Oph, V3890~Sgr does show a short interval of variability, mainly between days 15 and 18 (Figs.~\ref{lc} and \ref{zoom}) where, as frequently seen in these cases, the source is softer when brighter. The maximum change in count rate measured is a factor of $\sim$~6.5 in $\sim$~6~hr (between two {\em Swift} pointings) on day 15. In comparison, RS~Oph saw a change in brightness by a factor of $\sim$~13 in $\sim$~12~hr \citep{julo11}, while the X-ray count rate in V458~Vul increased by a maximum factor of $\sim$~40 in 1.5~hr \citep{ness09}.
As first described by \citet{julo11}, such hardness variations could be explained by clumpy absorption of the softer X-rays \citep[see also discussion in][]{schwarz11} although, as discussed in Section~\ref{spec}, the absorption column measured in V3890~Sgr appears to remain approximately constant after day 8.

Considering a shorter timescale, some novae also show quasi-periodic oscillations (QPOs) of a few tens of seconds during their SSS phase -- examples in {\em Swift} observations include RS~Oph, 35~s \citep{julo11, andy08, julo06}; KT~Eri, 35~s \citep{andy10}; V339~Del, 54~s \citep{andy13}; V5668~Sgr, 71~s \citep{page15a}. \citet{ness15} present a review of QPOs found in {\em XMM-Newton} and {\em Chandra} data, while the equivalent for all {\em Swift} data is in preparation (Beardmore et al.).

\begin{figure}
\begin{center}
  \includegraphics[clip, angle=-90, width=8.5cm]{Fig3a.ps}
  \includegraphics[clip, width=8.5cm]{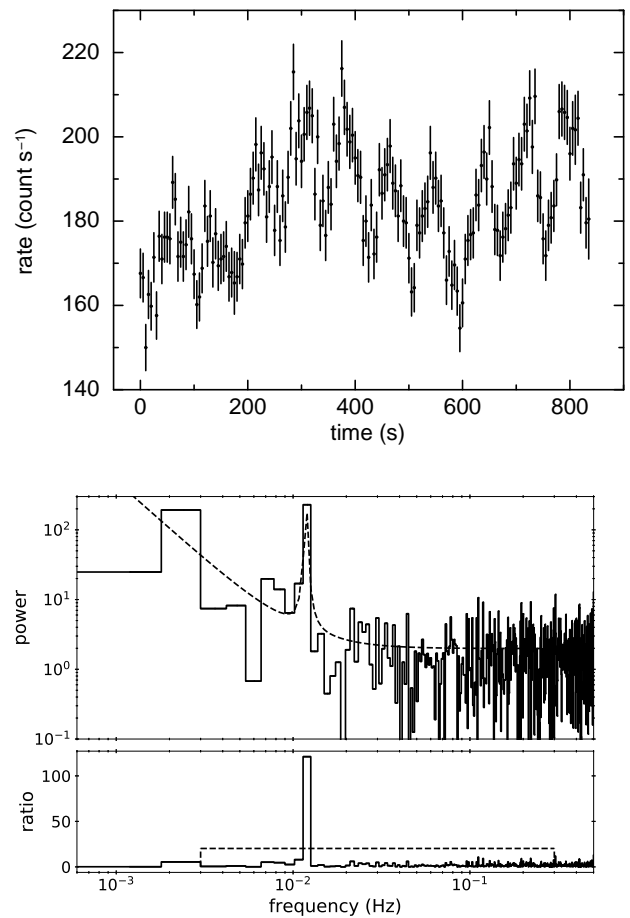}
\caption{Top: Soft band light-curve (0.3--0.85~keV) from the first {\em NICER} observation on 2019 September 06 (day 10), with 5~s bins. These data have not been background subtracted. Bottom: Periodogram analysis of the {\em NICER} data. The upper panel shows the periodogram (histogram) and fitted model (consisting of a powerlaw for
the low-frequency noise, a constant for the Poisson noise and a
Lorentzian for the oscillation). The lower window shows the ratio of
the periodogram to model when the Lorentzian is removed from the
model. The dashed line corresponds to the 99.73 per~cent (3$\sigma$) confidence
detection level over the frequency range 0.003--0.3~Hz.}
\label{qpo}
\end{center}
\end{figure}

While no QPO was evident in the {\em Swift} observations of V3890~Sgr, the first quick-look {\em NICER} dataset showed a clear detection of an 83~s oscillation \citep{beardmore19a}. As mentioned in Section~\ref{nicer} above, once the data were on the ground, it was found that the standard cleaning rejected this entire observation. Removing the usual filtering commands produced the light-curve shown in the top panel of Fig.~\ref{qpo}; even with the naked eye, a regular modulation is evident. We note that the SAA is not expected to vary quasi-periodically on a time frame of 83~s.

A detailed periodogram analysis of the {\em Swift}-XRT and {\em NICER} data was
performed to search for any sign of oscillations up to $\sim$~100~s.  This involved
examining periodograms from individual snapshots of (non-background-subtracted) data from each
instrument, as well as averaging the results to increase the
sensitivity of the search. We note that the background is negligible ($\sim$~0.5~per~cent) compared to the source emission at soft energies. The periodograms were normalised according
to \citet{leahy83}, which provides well-understood Poisson noise
properties when evaluating detection levels.  As continuous light
curves are required for a standard fast Fourier transform algorithm,
short gaps in the data were filled with the mean count rate before the
periodograms were computed.  For the XRT, only WT snapshots with mean
count rates above 1~count~s$^{-1}$ were considered.

The analysis revealed the detection of a significant modulation in the
first {\em NICER} dataset taken on 2019 September 06 (day 10), the light curve and
periodogram of which are shown in Fig.~\ref{qpo}, confirming the preliminary findings of \cite{beardmore19a}. The oscillation
period was calculated to be 82.9~$\pm$~0.6~s with an amplitude of 5.1~$\pm$~0.5~per~cent  \citep[using an epoch-folding period search technique; ][]{leahy87}.  The modulation is confined to the soft emission, though we cannot distinguish between it being caused by a change in blackbody (BB) normalisation, temperature, or both (see Section~\ref{spec} for spectral analysis).

The modulation was not seen in any of the other {\em NICER} observations (3$\sigma$ fractional amplitude upper limit of 2.1~per~cent), or their average (3$\sigma$ fractional amplitude upper limit of 1.1~per~cent). The modulation was
also not seen in the {\em Swift} data, to a 3$\sigma$ upper limit of 5.0~per~cent, averaged over 20 periodograms; unfortunately no {\em Swift} data were collected simultaneously with the {\em NICER} observations. There was also no QPO detected during the {\em XMM-Newton} observation around day 18 \citep{ness19b}.

It should be noted that the QPOs seen in other novae were not always visible, even during times of maximum SSS brightness, so the
non-detection in other {\em NICER} snapshots is not unprecedented.

\subsection{Spectral evolution}
\label{spec}

Preliminary analysis of the {\em Swift} X-ray spectra showed they were typically well-fitted by one or two optically-thin components \citep[{\sc xspec apec};][]{smith01} for the early- and late-time data. During the SSS phase, an additional BB component, further improved by including absorption edges, was required. 

\begin{figure}
\begin{center}
\includegraphics[clip, angle=-90, width=8cm]{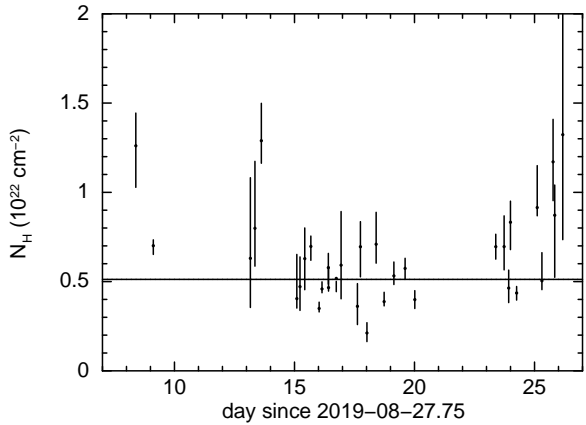}
\caption{Initial fits to the SSS spectra after day 8 show no evidence for a continued decrease in the absorbing column. The horizontal line in the plot marks the best fit constant value of 5.1~$\times$~10$^{21}$~cm$^{-2}$.}
\label{nh}
\end{center}
\end{figure}

As shown below, until the start of the SSS phase, the absorbing column was found to decrease steadily, as might be expected as the nova ejecta expand and thin, or as the shock traverses the RG wind \citep{bode06}; after this time, N$_{\rm H}$ was found to flatten off at about 5.1~$\times$~10$^{21}$~cm$^{-2}$, as demonstrated in Fig.~\ref{nh}. All fits after day 8 were therefore redone with N$_{\rm H}$ fixed at this value, to help constrain better the SSS parameters. The absorption and reddening of the system is discussed in more detail in Section~\ref{red}.

\subsubsection{Swift} 
The top panel of Fig.~\ref{lc} separates out the hard (2--10~keV) and soft (0.5--2~keV) X-ray light-curves, clearly demonstrating that the increase in brightness after day 8 is entirely due to soft counts below 2~keV.

\begin{figure}
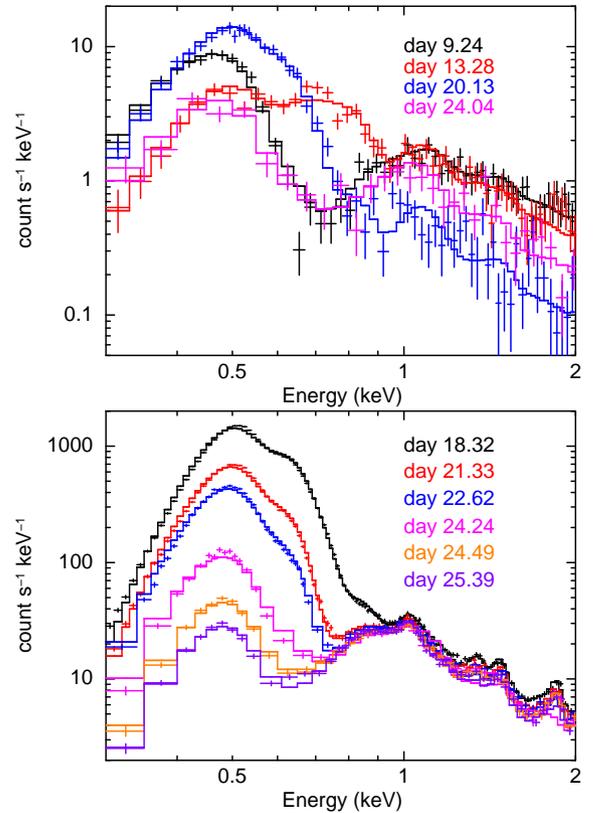

\begin{center}
\includegraphics[clip, angle=-90, width=8cm]{Fig5a.ps}
\includegraphics[clip, angle=-90, width=8cm]{Fig5b.ps}
\caption{Sample of X-ray spectra from {\em Swift} (top) and {\em NICER} (bottom), fitted with the model described in the text. The legend to the right provides the time since outburst when each spectrum was taken.}
\label{spec-evol}
\end{center}
\end{figure}

The upper panel of Fig.~\ref{spec-evol} shows a sample of the XRT SSS spectra, highlighting the increase in temperature between days 9.24 and 13.28, followed by a steady cooling/softening of the X-ray emission after this time. The spectra plotted were chosen to demonstrate the evolution of the soft emission clearly.

\begin{figure*}
\begin{center}
\includegraphics[clip, width=12cm]{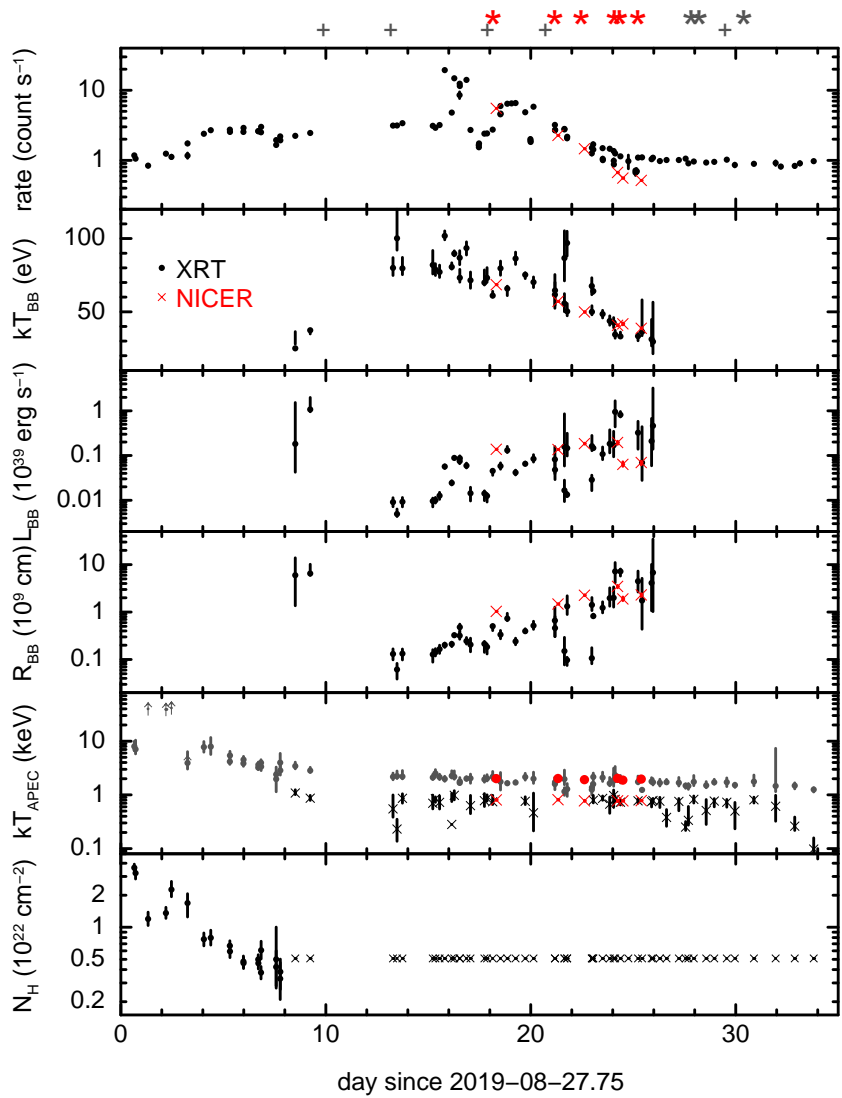}
\caption{Evolution of the X-ray spectra before, during and after the SSS phase.  Top panel: the 0.5--10~keV {\em Swift} XRT light-curve plotted as black circles; the red crosses show the 0.5-10 keV {\em NICER} count rate divided by 40 for comparison. Second, third and fourth panels: temperature, luminosity and radius from blackbody fits to the {\em Swift} and {\em NICER} data. Fifth panel: temperatures of the two {\sc apec} components from the {\em Swift} and {\em NICER} data, plotted as circles and crosses. Sixth panel: the cross symbols after day 8 indicate that N$_{\rm H}$ was fixed at 5.1~$\times$10$^{21}$ cm$^{-2}$ once the SSS component emerged. The stars and plus markers above the top panel mark the times of the clean {\em NICER} observations, and those where the data were affected by the SAA and/or high background (see text for details), respectively; the red stars highlight the times of the SSS spectra fitted. The error bars on the red {\em NICER} measurements are smaller than the size of the symbol. For reference, the Eddington luminosity is 1.26~$\times$~10$^{38}$~(M/$\msun$)~erg~s$^{-1}$.}
\label{spec-fit}
\end{center}
\end{figure*}

Fig.~\ref{spec-fit} shows the results of the spectral fits to the X-ray data from the first X-ray detection until 34 days later. The spectral fit parameters corresponding to this figure are provided in Table~\ref{swiftpar}.

\begin{table*}
  \caption{Fits to the {\em Swift} X-ray spectra before, during and after the supersoft emission. In addition to the parameters shown, absorption edges at 0.67, 0.74 and 0.87 keV are included when a BB is required (see main text for details). Where only a single {\sc apec} temperature is given, the cooler component was not statistically required. `Snap' indicates the ObsID consisted of more than one snapshot of data. The full version of this table is available online.}
  \begin{center}
    \begin{tabular}{lccccccc}
 
\hline
{\em Swift}  & Exposure & Day & BB kT & {\sc apec} kT$_{\rm hot}$ & {\sc apec} kT$_{\rm cool}$ & N$_{\rm H}$ & C-stat/dof\\
 ObsID   & time (ks) & & (eV) & (keV) & (keV) & (10$^{22}$ cm$^{-2}$) \\
 \hline

00045788003 (snap 1) & 1.33 &   0.6587 & &  7.99$^{+1.96}_{-1.35}$  & & 3.63$^{-0.26}_{+0.29}$ &  537/578\\
00045788003 (snap 2) & 0.66 &  0.7213	 & &  7.09$^{+3.37}_{-1.28}$  & & 3.23$^{-0.39}_{+0.34}$ & 333/402\\
00045788005 & 0.56  &        1.327 	 & &  $>$ 40.1 &  & 1.20$^{+0.19}_{-0.16}$ & 213/350\\
00045788006 & 0.65 &          2.199 	 & &   $>$ 39.1  & & 1.36$^{+0.18}_{-0.15}$ &  317/405\\
00045788007 & 0.28 &          2.467  &	 &   $>$ 41.6 & &  2.26$^{+0.44}_{-0.32}$ & 197/238\\

\hline
\end{tabular}

\label{swiftpar}
\end{center}

\end{table*}

\begin{figure}
\begin{center}
\includegraphics[clip, angle=-90, width=8cm]{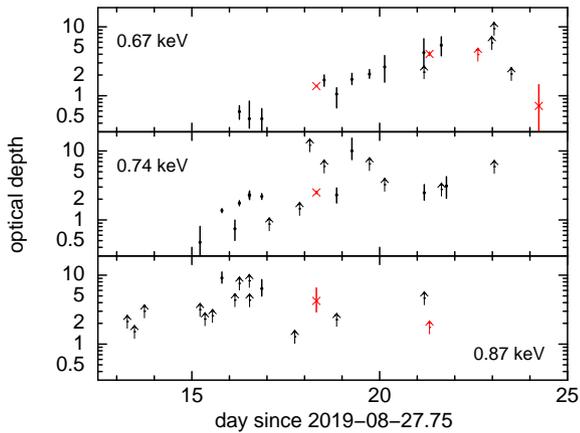}
\caption{Time evolution of the three absorption edges used to parametrize the XRT (black) and {\em NICER} (red) SSS spectra. Only the fits which improved the C-statistic by at least 6 for 1 degree of freedom (dof) have been plotted. Lower limit arrows are shown where the optical depth pegged at the allowed maximum of 50.}
\label{optdepth}
\end{center}
\end{figure}

From day 8.5, a BB component was included to model the new soft emission visible in the spectrum. Absorption edges with energies fixed at 0.67, 0.74 and 0.87 keV (H-like N, He-like O and H-like O) were included (these were only applied to the BB component; they did not affect the {\sc apec} components), with the optical depths allowed to vary freely (including down to zero if the edge was not required in the fit); Fig.~\ref{optdepth} demonstrates how the optical depths of these three edges vary with time. We caution that these edges are simply a way of improving the approximate BB parametrization of the soft emission (discussed in more detail in Section~\ref{fits}), with some of the fitted optical depths being poorly constrained, even when a significant improvement to the overall fit. However there is a trend that the 0.67~keV and 0.74~keV edges become deeper with time, until around day 24; for the highest energy edge at 0.87~keV, most of the optical depths were only lower limits, so nothing conclusive can be said about temporal evolution. After day 24, the BB temperature was too low ($\lo$~50~eV) for the edges to affect the component significantly.

The spectral fits were improved by a second optically-thin component after day 8.5, consistently around 0.5--1~keV; there was no significant evidence for this secondary temperature at earlier times, but this may be explained by the higher absorbing column quenching any such cooler emission. After day 26, the SSS component dropped below the 90~per~cent significance level, so the BB was no longer included, with the XRT spectra being well modelled by the two optically-thin components alone. There was an interval between days 20 and 23 where the second optically-thin component was not statistically required ($\lo$90~per~cent improvement). These data were collected in PC mode, while the spectra before and after were taken using WT mode; because the count rate was still sufficiently high for the PC data to be piled-up, annular extraction regions were used, leading to the spectra being of lower statistical quality than the others. This is likely to be the reason why the cooler {\sc apec} was not significantly required, given that the {\em NICER} spectra during this interval required both temperatures.
We also note that the \emph{Chandra} grating spectra collected around day 6--7 were better fitted by a dual temperature model with kT~$\sim$~1 and 4~keV \citep{orio20}, in agreement with our results here.

At early times, the observed 0.5--10~keV flux of the hotter optically-thin plasma varies between 0.7--1.4~$\times$~10$^{-10}$~erg~cm$^{-2}$~s$^{-1}$; after around day 4, the flux fades steadily, decreasing by an order of magnitude, from $\sim$~1.4~$\times$~10$^{-10}$~erg~cm$^{-2}$~s$^{-1}$ to $\sim$~1.5~$\times$~10$^{-11}$~erg~cm$^{-2}$~s$^{-1}$ by day 33. 
The cooler component is fainter, but appears to fade at about the same fractional rate as the hotter component during the interval they are both detectable (they both decrease in flux by a factor of $\sim$2--3 between days 15 and 33), declining to $\sim$~5~$\times$~10$^{-12}$~erg~cm$^{-2}$~s$^{-1}$ by the time of the final spectrum fitted. 

\subsubsection{NICER}
\label{nicerspec}

Following the {\em Swift} analysis, a model consisting of two {\sc xspec vapec} optically-thin components, together with a BB component and three absorption edges for the SSS emission was the baseline model from which we started the {\em NICER} spectral fitting. 
With many more counts in the {\em NICER} spectra (useful energy range of 0.2--12~keV; however, the spectra for V3890~Sgr become background-dominated above 5~keV, so the higher energies were excluded), they can be used to place limits on abundances which cannot be easily done using the {\em Swift}-XRT data. While grating instruments are more desirable for detailed abundance investigations, only a single observation of V3890~Sgr was taken by each of {\em Chandra} \citep{orio19, orio20} and {\em XMM-Newton} \citep{ness19b} during the outburst. 

\citet{orio20} found that the {\em Chandra} High Energy Transmission Grating spectrum obtained on the seventh day after outburst could be fitted with two optically-thin plasma components, with the cooler one showing evidence for enhanced metal abundances. With this in mind, and after some experimentation with the {\em NICER} data, the abundances of oxygen, neon, magnesium and silicon in the cooler component were allowed to vary, while all others remained fixed at solar. This decreased the residuals in the fits without significantly altering the measured temperatures. In agreement with {\em Chandra}, neon, magnesium and silicon were all found to be over-abundant in all fits ($\sim$~3--6 times solar), while the oxygen abundance was initially consistent with zero, before apparently becoming super-solar at later times. Varying these abundances in the {\em Swift} fits made no significant differences to those results. Table~\ref{nicerabund} lists the parameters derived from the fits to the {\em NICER} data obtained during the SSS phase. Although we model the harder X-ray emission with just two different temperature components, this is simply a parametrization of the more complex underlying continuum shock emission \citep[c.f., ][]{vaytet07, vaytet11}. For comparison, \cite{steve11} present the 2010 eruption of the similar symbiotic RN V407~Cyg, analysing the shock and its evolution in that similar system in detail.

\begin{table*}

\caption{Fits to the {\em NICER} SSS spectra. Only the abundances listed were allowed to vary, all others remained at the solar value; see text for discussion. In addition to the components shown, three absorption edges at 0.67, 0.74 and 0.87 keV were fitted.}

\begin{center}
\begin{tabular}{lcccccccccc}
\hline
    {\em NICER}  & Exposure & Day & BB kT & {\sc apec} kT$_{\rm hot}$ & {\sc vapec} kT$_{\rm cool}$ & \multicolumn{4}{c}{Abundances} & C-stat/dof\\
 ObsID   & time (ks) & & (eV) & (keV) & (keV) & O & Ne & Mg & Si\\
\hline
2200810104 & 1.87 & 18.32 & 68.5~$\pm$~0.1 & 2.01$^{+0.05}_{-0.04}$ & 0.81~$\pm$~0.2 & $<$0.33 & 5.7~$\pm$~0.6 & 4.1$^{+0.4}_{-0.3}$ & 3.3$^{+0.3}_{-0.2}$  &1902/455\\
2200810106 & 1.51 & 21.33 & 57.1 ~$\pm$~0.1 & 2.02$^{+0.05}_{-0.04}$ & 0.82~$\pm$~0.2 &$<$0.16 & 6.5$^{+0.5}_{-0.6}$ & 3.7~$\pm$~0.2 & 3.1~$\pm$~0.2 & 1117/453 \\
2200810107 & 0.72 & 22.62 & 49.2~$\pm$~0.2 & 1.92~$\pm$~0.07 & 0.78~$\pm$~0.2 &  $<$0.24 & 4.4$^{+0.6}_{-0.7}$ & 3.2~$\pm$~0.3 & 2.7$^{+0.3}_{-0.2}$  &701/425\\
2200810108 & 0.41 & 24.24 & 40.7$^{+0.5}_{-0.3}$ & 2.03$^{+0.11}_{-0.10}$ & 0.78~$\pm$~0.2 & $<$0.28 & 5.1$^{+0.9}_{-0.8}$ & 3.7$^{+0.4}_{-0.3}$ & 3.1$^{+0.4}_{-0.3}$ &443/362  \\
2200810109 & 0.68 & 24.49 & 41.8$^{+0.6}_{-0.7}$ & 1.88$^{+0.08}_{-0.07}$ & 0.78$^{+0.01}_{-0.02}$ & $<$1.1 & 4.6$^{+0.7}_{-0.6}$ & 2.9~$\pm$~0.3 &2.8~$\pm$~0.3 & 421/404\\
2200810110 & 1.13 & 25.39 & 38.8$^{+0.3}_{-0.4}$ & 1.97$^{+0.07}_{-0.08}$ & 0.78~$\pm$~0.01 & 4.6$^{+0.8}_{-0.5}$ & 5.7 $^{+0.4}_{-0.3}$ & 2.9~$\pm$~0.2 & 2.9~$\pm$~0.2 & 422/434\\

\hline
\end{tabular}

\label{nicerabund}
\end{center}
\end{table*}

The {\em NICER} spectra are plotted in the lower panel of Fig.~\ref{spec-evol}, and the BB temperatures, luminosities and effective radii measured for the SSS component included in Fig.~\ref{spec-fit}, showing them to be in good agreement with the {\em Swift} results.

None of the {\em NICER} observations was strictly simultaneous with {\em Swift}, so, given the obvious spectral variability, joint fits were not performed.

\subsubsection{Alternative spectral fits}
\label{fits}

\begin{figure*}
\begin{center}
\includegraphics[clip, angle=-90, width=15cm]{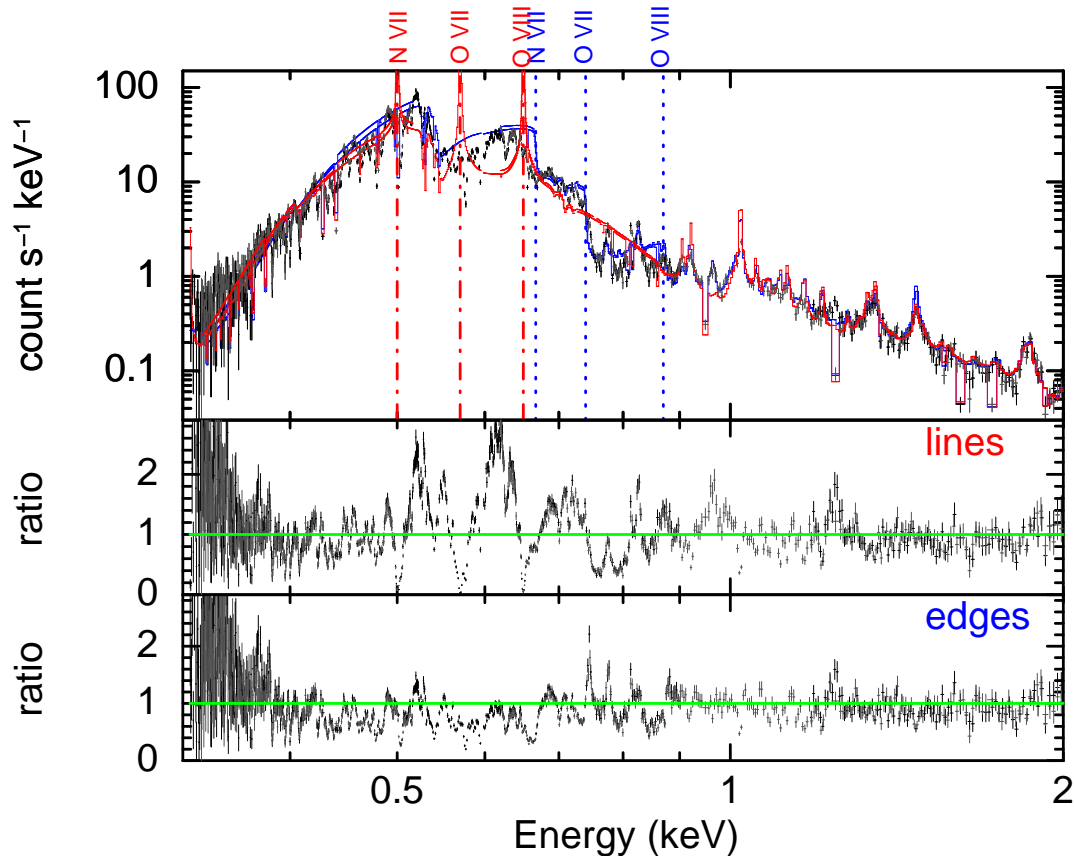}
\caption{{\em XMM-Newton} RGS spectra (ObsID 0821560201) overlaid with models from the {\em NICER} ObsID 2200810104 dataset. The top panel shows the model consisting of a BB and three emission lines (as well as two underlying optically-thin components) in red (line energies marked in red with dot-dashed lines), while the blue solid line shows the model with absorption edges instead (edge energies marked in blue with dotted lines). Panels 2 and 3 show the respective residuals.}
\label{edgelinecomp}
\end{center}
\end{figure*}

While the addition of three absorption edges to the underlying BB model is one way to parametrize the data, an equally good fit can be obtained using a BB together with three narrow emission lines at 0.50, 0.57 and 0.65~keV, corresponding to H-like N, He-like O and H-like O respectively, when considering the first, brightest, {\em NICER} dataset, for example\footnote{We note this is not necessarily a physically-based model, but simply an example to demonstrate that the addition of absorption edges is not a uniquely acceptable model.}. While CCD spectra such as those from {\em Swift}-XRT and {\em NICER} may not be able to differentiate between such models, grating spectra are better suited to this. An observation was obtained with the {\em XMM-Newton} Reflection Grating Spectrometer \citep[RGS; ][]{rgs} around 18~days after the eruption \citep[ObsID 0821560201; ][]{ness19b}, overlapping {\em NICER} observation 2200810104. The best fits to the {\em NICER} data, using a BB together with absorption edges or emission lines, were folded through the RGS response and over-plotted on those data. The results, shown in Fig.~\ref{edgelinecomp}, clearly demonstrate that the fit with the emission lines is inconsistent with the RGS data, with the edge model being in much better agreement; while we do not claim these models to be directly appropriate for the grating spectra, the Cash-statistic results are 63649 and 141703 for 5377 dof for the edge and line models respectively, showing the edge fit to be far better. As an aside, \cite{ness13} find that SSS spectra which show emission lines tend to be obscured systems, where the continuum is suppressed. The fit to the RGS data can be further improved by increasing the oxygen abundance in the absorption component, but that is beyond the scope of this paper; the RGS data will be presented in full by Ness et al. (in prep.).

Modelling the SSS component with a BB alone, BB plus edges or BB plus emission lines all lead to different fitted BB temperatures: absorption edges tend to increase the BB kT, while emission lines decrease the value. That is, the precise parameter values of the super-soft emission depend on the model used. While a simple BB is not a statistically good fit to the vast majority of the SSS spectra presented here, we did confirm that the underlying trends (for example, when, and the rate at which, the temperature decreases) were the same for this more basic model. Thus the evolution of the SSS emission is approximately model-independent.

While it is known that BB models are an over-simplification of the emission from the WD, and are not physically realistic \citep[having the potential to underestimate the temperature and overestimate the luminosities; see, e.g.,][although \citealt{julo11} found consistent results from the BB and atmosphere fits during the SSS plateau phase of RS~Oph]{kraut96}, the available model atmosphere grids\footnote{We considered the TMAP T{\" u}bingen NLTE Model Atmosphere Package of plane-parellel, static, non-local-thermal-equilibrium models: http://astro.uni-tuebingen.de/\raisebox{.2em}{\tiny$\sim$}rauch/TMAF/flux\_HHeCNONeMgSiS\_gen.html} did not reach high enough temperatures to fit the V3890~Sgr data, with many of the spectral fits pegging at uppermost temperature bound of 1.05~$\times$~10$^{6}$~K (90.5~eV). The inclusion of the absorption edges in addition to the BBs does, however, somewhat mitigate the issue. Considering a sample of M31 novae, \citet{hen11} found a strong correlation between the temperatures estimated using BB fits and those using model atmospheres, suggesting that BBs can be used to parametrize the SSS temperature trends.  In addition, high-resolution X-ray spectra have shown complex results (see Fig.~\ref{edgelinecomp}), highlighting possible complications with using idealised atmosphere models and indicating more phenomological approaches may be advisable \citep{ness19a}. 

\section{UV Spectra}

It was found that the early UVOT grism spectra (before the observations were
offset from day 16) suffered considerably from zeroth order
contamination from field stars, while for some roll angles the first
order spectrum of a field star overlaid the nova spectrum; for the
affected parts of those spectra, only the emission lines are useful.

\subsection{UV continuum}
\label{cont}

The UVOT grism spectra span 1700--4500~\AA, covering the Balmer jump at 3646~\AA; Fig.~\ref{grism} plots the spectra which were not contaminated by first order emission from field stars. In a normal stellar atmosphere, the flux shortward of the Balmer jump is smaller, due to the larger opacity caused by bound-free absorption transitions. However in our UV spectra, the continuum at shorter wavelengths is actually brighter. This can occur in the highly extended atmospheres of novae, which can have a larger photosphere at the higher opacities below the jump than above it \citep{geb68, harkness83, hauschildt92}. Using the data presented here, we observe that the strength of the Balmer jump increases over time between days 6 and 18, the reason being that the continuum is formed in the extended medium of the shocked RG wind and the nova ejecta.

\subsubsection{Comparison with IUE}

\begin{figure}
\begin{center}
\includegraphics[clip, width=9cm]{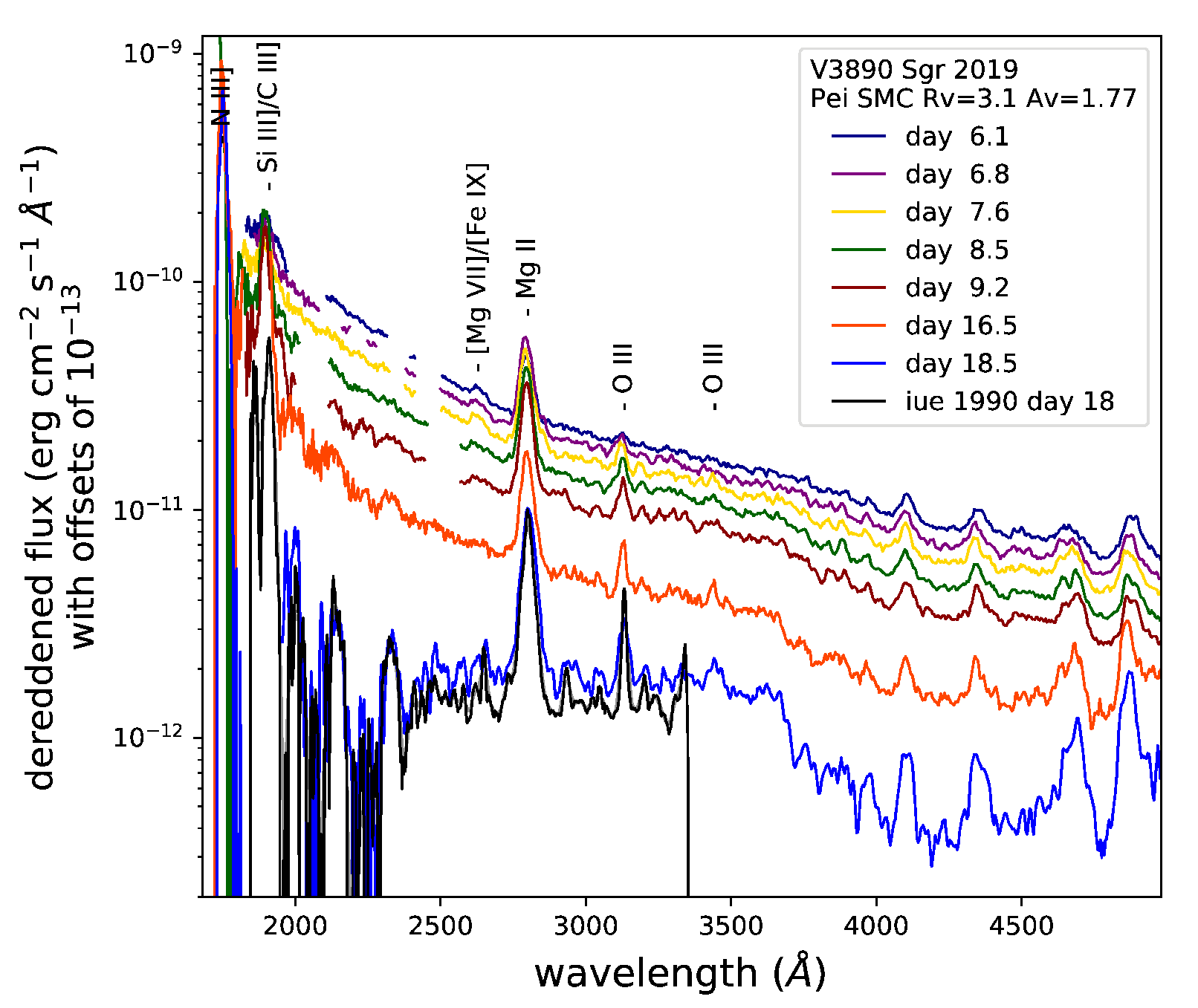}
\caption{Sample of {\em Swift} UVOT grism spectra of V3890 Sgr dereddened using an SMC reddening law. The spectra have been vertically offset by 10$^{-13}$ erg~cm$^{-2}$~s$^{-1}$~\AA$^{-1}$ for clarity, and those after day 18 have been smoothed using a 3-point boxcar algorithm. Lines mentioned in Section~\ref{sec:lines} are marked. The IUE spectrum from day 18 after the 1990 eruption has been included for comparison.}
\label{grism}
\end{center}
\end{figure}

Data were obtained by {\em IUE} ({\em International Ultraviolet Explorer}) on days 18 and 26 after the previous 1990 eruption of V3890~Sgr \citep{gonz92}; the first of these spectra is included in Fig.~\ref{grism}. The long wavelength {\em IUE} data closely match the UVOT spectrum on day 18 after the recent eruption, though with a slightly lower flux and narrower lines, suggesting the start date for the 1990 event was slightly earlier than that adopted by \citet{gonz92}. The day 18 spectra from both 1990 and 2019 show weak lines at 2143~\AA\ and 2332~\AA, corresponding to N\,II and [O II] respectively, with a full width at zero intensity (FWZI) of 10,000~km~s$^{-1}$.

\subsection{UV line emission}
\label{sec:lines}

The UVOT spectrum shows UV emission lines of N\,III] 1750~\AA, the blend of Si\,III] 1892\AA/C\,III] 1909~\AA, a broad line at 2630~\AA\ which is a [Mg\,VII] blend with weaker [Fe\,IX] at 2648~\AA, Mg\,II 2800~\AA, and the Bowen lines of O\,III at 3133~\AA\, and 3444~\AA. Similar line identifications were made by \cite{gonz92} for the 1990 eruption. No lines of [Ne\,III] are seen at 3870/3970~\AA, nor is there a [Ne\,V] doublet at 3349/3426~\AA, however the upper limits on the Ne abundance are close to solar.

\begin{figure}
\begin{center}
\includegraphics[clip, width=9cm]{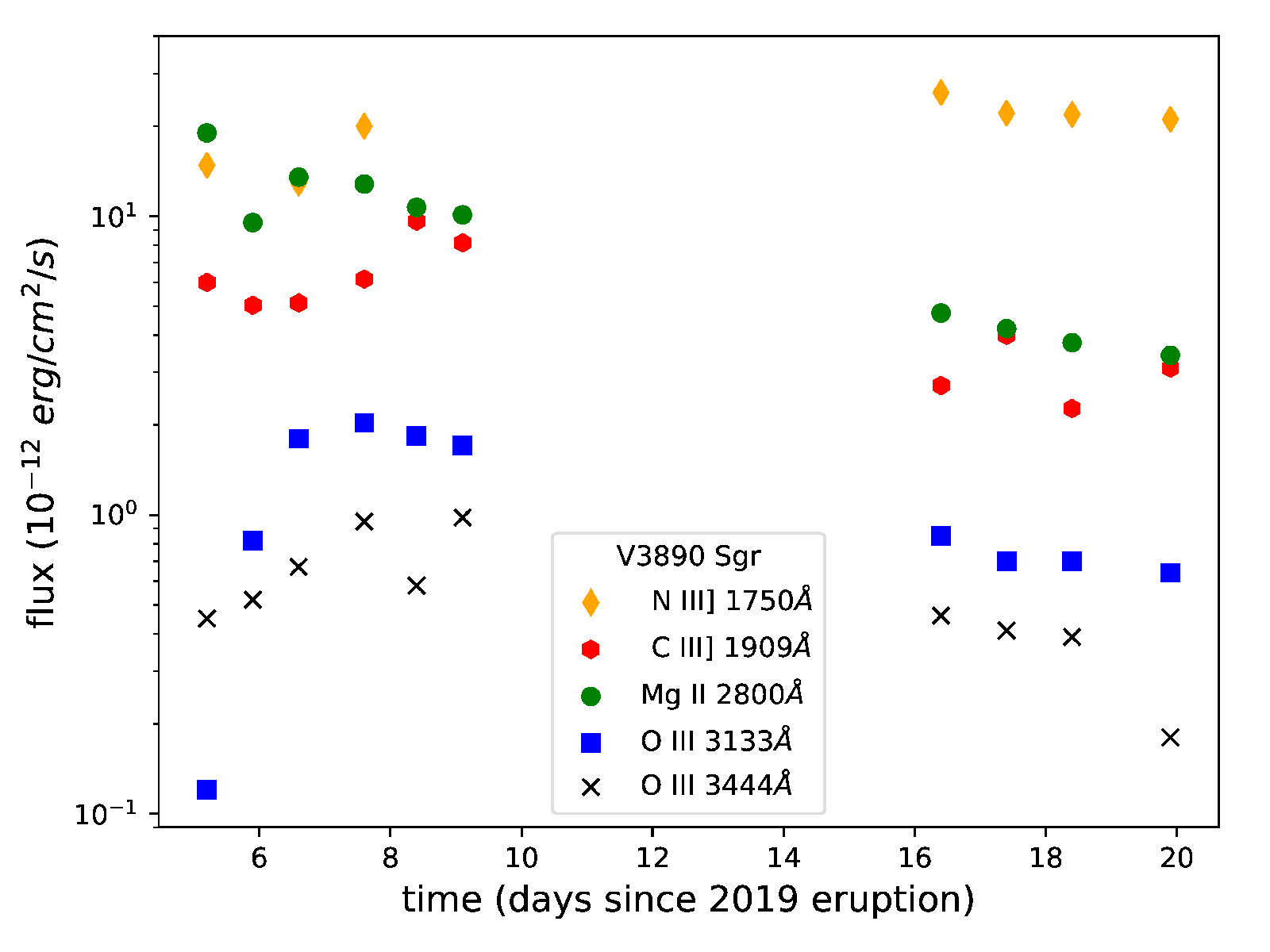}
\caption{Evolution of the UV spectral line flux. Errors have not been shown but are estimated to be 15~per~cent, except in N\,III], where the uncertainty is typically 30~per~cent.}
\label{uvlines}
\end{center}
 \end{figure}

      Figure~\ref{uvlines} shows the evolution of the line fluxes. The Mg\,II resonance line shows a steady decrease in flux, while the O\,III 3133~\AA\ Bowen line is seen to gain in strength until day~7, likely because the ejecta were still optically thick to the He Ly$\alpha$ 304~\AA\ line which pumps 
the O\,III emission at early times. Due to the noise in the spectra, the width of the Mg\,II line cannot be said to change significantly, with FWZI/2~=~7500~$\pm$~1000~km~s$^{-1}$, and 
full width at half maximum FWHM~=~5000~$\pm$~500~km~s$^{-1}$. The core of the line profile becomes narrower at later times, while the wings remain extended; the line widths (FWZI/2 $\sim$~55~\AA\, at 2800~\AA) are significantly larger than the FWHM of the Point Spread Function (PSF; $\sim$~3~\AA). Flux errors of $\sim$~15~per~cent ($\sim$~30~per~cent for N\,III]) are estimated from the differences seen in spectra obtained on the same day, before they were averaged.

\subsection{Reddening}
\label{red}

Using the {\em IUE} spectra from the 1990 eruption of V3890~Sgr, \citet{gonz92} estimated a value of E(B$-$V)~=~1.1~$\pm$~0.1, while \citet{brad10} lists an average extinction for the system of E(B$-$V)~=~0.9. Using data from the current eruption, \citet{mun19} derive E(B$-$V) to be somewhat lower, between 0.56 and 0.62, with a mean value of 0.59 mag. The reddening derived from the IRAS/COBE-DIRBE 100~$\mu$m maps \citep[]{sch98}\footnote{These maps provide the {\em total} Galactic reddening in any given direction, not just to the distance of the nova.}, and updated using Sloan spectra \citep{sch11}, yields E(B$-$V)~=~0.4822~$\pm$~0.0091. By combining the earlier dust maps with Pan-STARRS PS1, the 2MASS photometry and distances from {\em Gaia} DR2, \citet{green19} derive 3-D reddening maps.
They define an extinction parameter, E, for the assumed mean Galactic reddening law scaled to the \citet{sch98} results. In the case of V3890~Sgr, E~=~0.561, which is valid out to a distance of $\sim$~4~kpc along the line of sight. 
Their conversion to A$_{\rm V}$ has been based on a comparison to APOGEE DR14 \citep[Apache Point Observatory Galactic Evolution Experiment; ][]{queiroz18}, providing a value of A$_{\rm V}$~=~1.77~$\pm$~0.09~mag, which we adopt here.

Initally, we tried to apply a reddening correction to the {\em Swift} UV spectrum using the \citet{ccm89} Galactic extinction curve for various combinations of 
E(B$-$V) and R$_{\rm V}$ giving a total visual extinction of A$_{\rm V}$~=~1.77 mag. However, any reasonable value of R$_{\rm V}$ led to a pronounced bump in the dereddened spectrum at 2175~\AA. Therefore, different reddening laws were investigated, using the parametrization by \citet{pei92} of the Galactic (Milky Way; MW), Large Magellanic Cloud (LMC) and Small Magellanic Cloud (SMC) extinction curves, where notably the effect of the 2175~\AA\ absorption peak decreases from MW to LMC to SMC. An LMC reddening law with R$_{\rm V}$~=~3.1 and A$_{\rm V}$~=~1.77 mag was not favoured since it leads
to a small bump at 2200~\AA. However, we do find a smooth corrected spectrum using the SMC curve; see Fig.~\ref{grism}. We note that is not unusual to find deviations from the average MW extinction law, and that there are differences in the 2175~\AA\ bump even within a galaxy \citep[e.g., ][]{lea17}.
The derived intrinsic UV flux is high since the SMC extinction law is steeper
in the UV than the mean Galactic reddening.

The slope of the UV spectrum indicates a fairly hot emission source, with
equivalent BB temperatures of the order of 10$^5$~K ($\sim$~10~eV). The V3890~Sgr symbiotic system, similar to RS~Oph and V407~Cyg \citep{julo11, steve11, steve12}, comprises a WD and a RG. The evolved RG has an extensive wind, with typical densities in the range of 10$^{4-9}$~cm$^{-3}$, velocities of 10s--100s km~s$^{-1}$ and temperatures of up to 10$^{4-5}$~K \citep[in part due to shocks by the ejecta; ][]{reimers77, steve11}. The initial nova explosion flash ionizes the wind, which will then experience a strong mechanical shock from the ejecta interacting with it. Comparison with the detailed spectral component analysis of V407 Cyg by \cite{steve11,steve12} suggests that the source of the UV continuum is dominated by the photoionized RG wind.


The Leiden/Argentine/Bonn (LAB) and Galactic All-Sky Survey (GASS) measurements provide an estimate of N$_{\rm H}$~=~$\sim$~1.9~$\times$~10$^{21}$~cm$^{-2}$ \citep{kal05, kal15} in the direction of V3890~Sgr, which is lower than the value derived from the fits to the XRT spectra presented here (5.1~$\times$~10$^{21}$~cm$^{-2}$) by a value of $\sim$~3.2~$\times$~10$^{21}$~cm$^{-2}$. The column including molecular hydrogen \citep[N$_{\rm H,tot}$~=~N$_{\rm H_I}$~+~2N$_{\rm H_2}$;][]{dick13}\footnote{\url{https://www.swift.ac.uk/analysis/nhtot/}} is $\sim$~3~$\times$~10$^{21}$~cm$^{-2}$, still below the best fit from the XRT data. Using the relation between optical extinction and N$_{\rm H}$ found by \citet{go09} of N$_{\rm H}$ (cm$^{-2}$)~=~(2.21~$\pm$~0.09)~$\times$~10$^{21}$~A$_{\rm V}$ (mag), our adopted value of A$_{\rm V}$~=~1.77 suggests a column of 3.9~$\times$~10$^{21}$~cm$^{-2}$.
With a back-of-the-envelope calculation, we can explain this discrepancy as being due to additional absorption in the wind of the RG star. The luminosity of the secondary was estimated using the SED based on published photometry taken from the Vizier photometry tool\footnote{\url{http://vizier.unistra.fr/vizier/sed/}}, with the spectrum below 5000~\AA\,
replaced by a scaled spectral model for an M5 III star \citep{pickles98} and effective temperature consistent with the spectral type. Lines in the Chiron spectra (obtained in 2018, between nova eruptions) from the Stony Brook/SMARTS Atlas of (mostly) Southern Novae\footnote{\url{http://www.astro.sunysb.edu/fwalter/SMARTS/NovaAtlas/atlas.html}} \citep{fred12} were examined, identifying the low density forbidden line width of [O\,I] as giving the final wind velocity, while the lines formed at higher densities are indicative of the turbulent velocities in the subsonic region. The mass loss rate was obtained using the modified \citet{reimers77} formula from \citet{sch05}, with the density in the wind following; this can then be related to the column density at a distance of 1.3--10 stellar radii, providing an estimate of the order of the excess N$_{\rm H}$ seen in the XRT spectra (i.e. an additional $\sim$~(1--3)~$\times$~10$^{21}$~cm$^{-2}$ above the interstellar value required to reach the fitted column of $\sim$~5.1~$\times$~10$^{21}$~cm$^{-2}$, depending on the method of estimating the Galactic value). Seeing excess absorption from the RG wind suggests that the nova eruption occurred when the WD was on the far side of the binary system from our view point.

It has previously been considered that clumpy absorption could be the cause of the high-amplitude flux variability seen during the onset of the SSS phase in some novae. However, there was no strong evidence for changing N$_{\rm H}$ after day 8 in the case of V3890~Sgr.

\section{Optical spectra}

\begin{table}

\caption{Optical lines seen in the spectra. The first column gives the line identification, its wavelength in \AA\, and, where relevant, the multiplet in parentheses; the second column lists the day range during which the line was visible, with the interval where the flux is $>$5~per~cent of the peak value given in parentheses; the third column lists the measured peak flux and the fourth column gives the day of the peak.}

\begin{center}
\begin{tabular}{lccc}
\hline
Line  & Days visible & Peak flux & Day of \\
& (Days visible at & (erg~cm$^{-2}$ & peak\\
& $>$5~per~cent peak flux)& s$^{-1}$~\AA$^{-1}$)\\
\hline
N\,{\sc iii}      4640  & 3.27 -- 41.25  &     2.6 $\times$~10$^{-12}$ &   7.28\\
& (5.34 -- 22.27)\\
He\,{\sc ii}      4686 &   3.27 -- 45.23  &      9.7 $\times$~10$^{-12}$ &   8.27\\
& (5.34 -- 23.29)\\
He\,{\sc i}      4713 &   3.27 -- 29.26  &      1.1 $\times$~10$^{-12}$ &   5.34\\
&  (3.27 -- 21.27)\\
H\,{\sc i}        4861 &   3.27 -- 49.23 &       5.4 $\times$~10$^{-11}$ &   3.27\\
& (3.27 -- 23.29)\\
He\,{\sc i}   4921 &   1.31 -- 45.23  &      3.2 $\times$~10$^{-12}$ &   7.28\\
& (1.31 -- 27.33)\\
Fe\,{\sc ii} (49) 5197 &   1.31 -- 35.29 &       5.5 $\times$~10$^{-13}$ &   8.27\\
& (1.31 -- 29.26)\\
Fe\,{\sc ii} (49) 5234 &   1.31 -- 61.24 &       9.0 $\times$~10$^{-13}$ &   2.33\\
& (1.31 -- 27.33)\\
Fe\,{\sc ii} (49) 5276 &   1.31 -- 41.25 &      1.7 $\times$~10$^{-12}$ &   1.31\\
& (1.31 -- 23.29)\\
Fe\,{\sc ii} (41) 5284 &   1.31 -- 45.23  &      8.5 $\times$~10$^{-13}$ &   2.33\\
& (1.31 -- 25.33)\\
$[$Fe\,{\sc xiv}]   5303 &   5.34 -- 35.29 &     1.8 $\times$~10$^{-12}$ &  18.26\\
& (15.28 -- 19.27)\\
Fe\,{\sc ii} (49) 5316 &   1.31 -- 49.23 &       3.2 $\times$~10$^{-12}$ &   1.31\\
& (1.31 -- 25.33)\\
$[$Fe\,{\sc vii}]   5721 &   5.34 -- 38.28 &      5.6 $\times$~10$^{-14}$ &   7.28\\
& (5.34 -- 27.33)\\
S\,{\sc ii}       6084 &   1.31 -- 35.29 &       6.9 $\times$~10$^{-14}$ &  15.28\\
& (1.31 -- 28.26)\\
$[$Fe\,{\sc vii}]   6087  &  1.31 -- 53.26 &      1.2 $\times$~10$^{-13}$ &   7.28\\
& (1.31 -- 45.23)\\
Si\,{\sc ii}   6347 &   1.31 -- 45.23 &       8.0 $\times$~10$^{-13}$ &   8.27\\
& (3.27 -- 25.33)\\
$[$Fe\,{\sc x}]     6374 &   1.31 -- 61.24 &       2.7 $\times$~10$^{-12}$ &  21.27\\
& (2.33 -- 45.23)\\
Fe\,{\sc ii}      6383 &   1.31 -- 53.26 &       4.2 $\times$~10$^{-13}$ &  11.31\\
& (1.31 -- 38.28)\\
Fe\,{\sc ii} (74) 6416 &   1.31 -- 53.26  &     2.5 $\times$~10$^{-13}$ &   8.27\\
& (1.31 -- 35.29)\\
H\,{\sc i}       6563 &   1.31 -- 61.24 &      5.9 $\times$~10$^{-10}$ &   2.33\\
& (1.31 -- 27.33)\\
He\,{\sc i}       7065 &   1.31 -- 61.24  &     3.5 $\times$~10$^{-11}$ &   1.31\\
&(1.31 -- 24.38)\\
O\,{\sc i}        7774 &  11.31 -- 24.38  &    2.0 $\times$~10$^{-12}$ &  11.31\\
&(11.31 -- 20.23)\\ 
Mg\,{\sc ii}      7786 &   2.33 -- 27.33  &      7.7 $\times$~10$^{-11}$ &  2.33\\
&(2.33 --  3.27)\\
$[$Fe\,{\sc xi}]    7891 &   4.30 -- 29.26  &   2.0 $\times$~10$^{-12}$ & 23.29\\
&(11.31 -- 23.29)\\

\hline
\end{tabular}

\label{smarts}
\end{center}
\end{table}

\begin{figure*}
\begin{center}
  \includegraphics[clip, angle=-90, width=15cm]{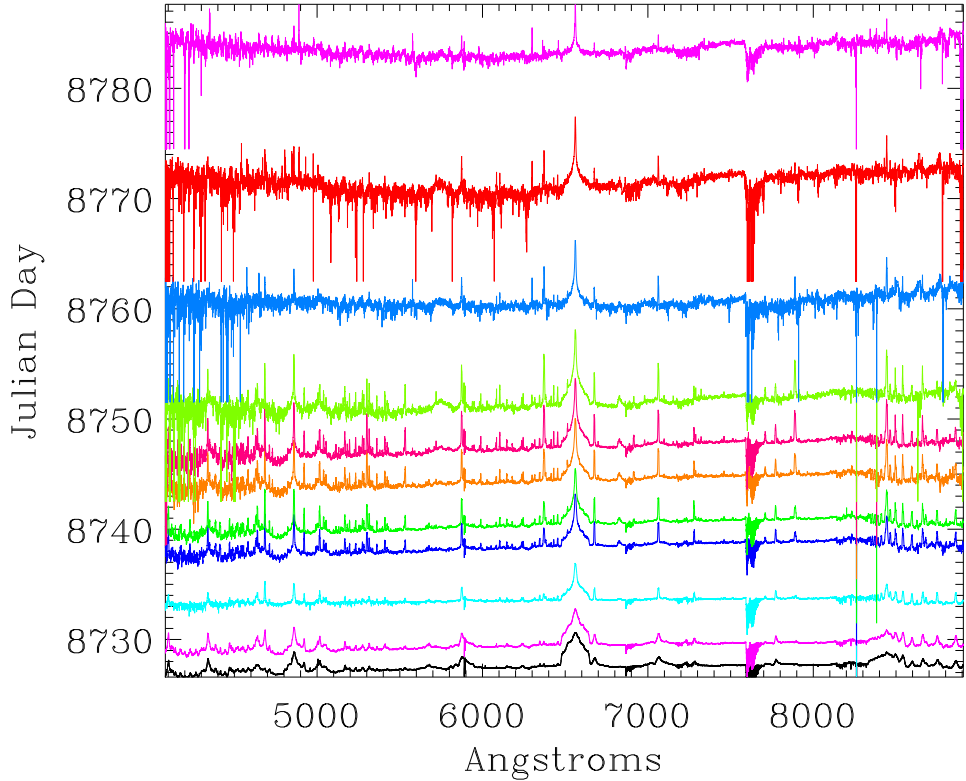}
\caption{Sample of some of the Chiron spectra we used in this analysis. The scaling is logarithmic to show lines weaker than H$\alpha$. 
Continua are offset to the Julian date (245+) of the observation.
The continuum is noisy in the later spectra as the nova fades, especially
shortward of 4700\AA. The data are smoothed, and points less than 1~per~cent of
the median continuum flux have been truncated at that level.} 
\label{smartsspec}
\end{center}
\end{figure*}

\begin{figure*}
\begin{center}
  \includegraphics[clip, angle=-90, width=8.5cm]{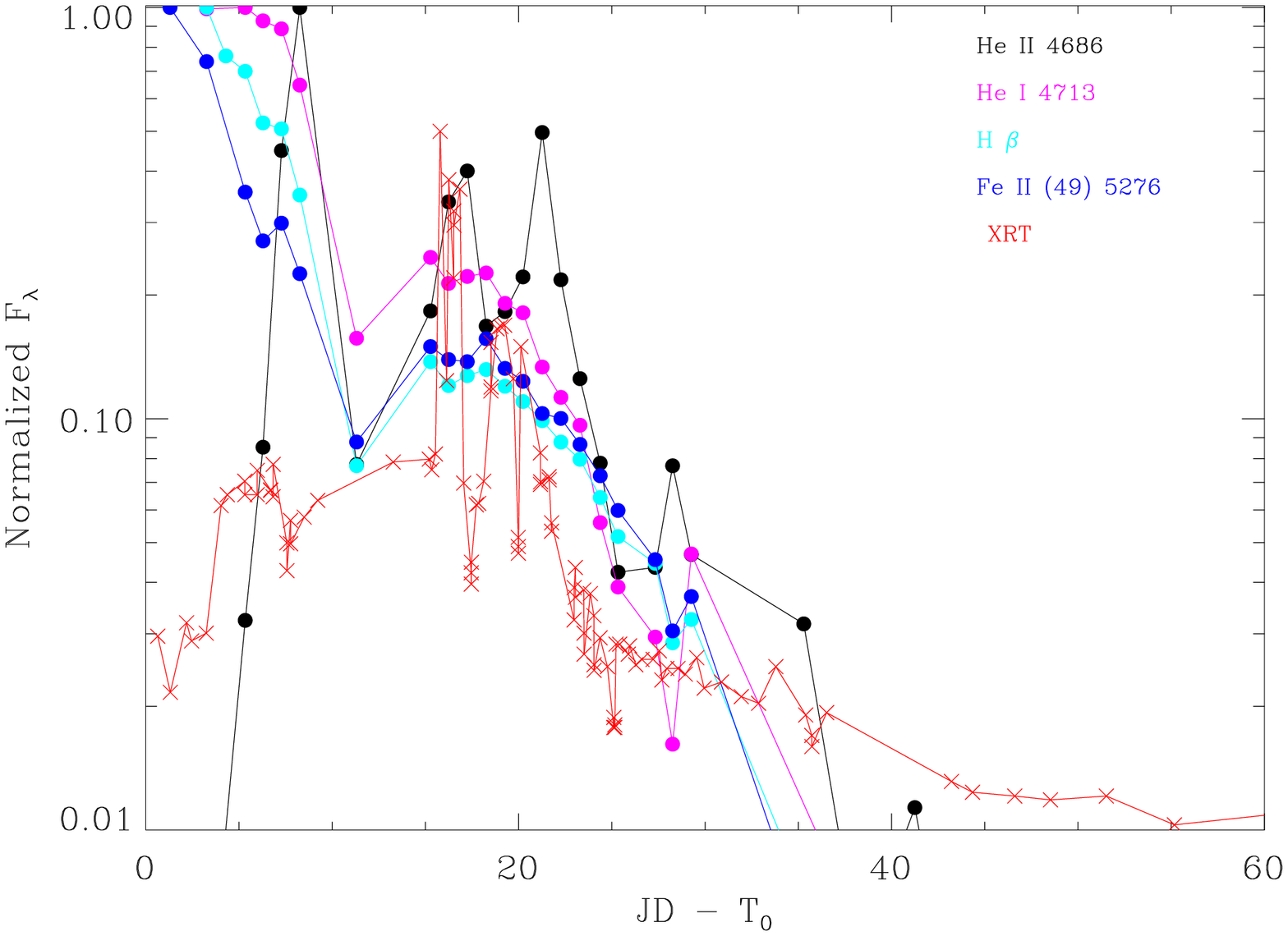}
  \includegraphics[clip, angle=-90, width=8.5cm]{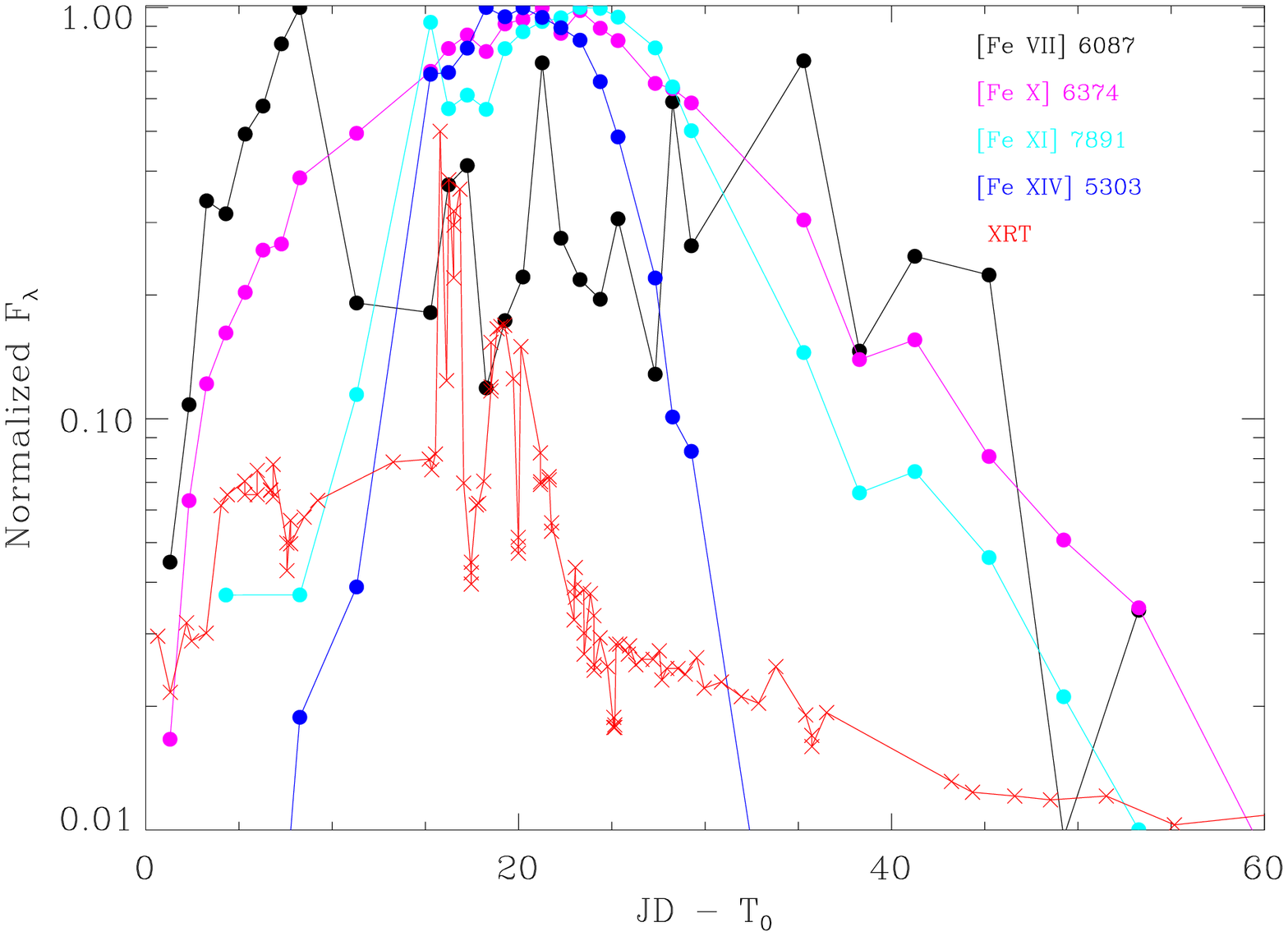}
\caption{The time variation of four permitted (left panel) and four forbidden (right panel) lines compared to the X-ray light-curve (shown in red). The lines are scaled to their respective maxima, while the X-rays have been reduced by half for clarity. The permitted Fe\,{\sc ii}, He\,{\sc i} and H$\beta$ lines all peak close to the time of the nova, before the appearance of the SSS,
and have a secondary peak following the X-ray maximum. He\,{\sc ii} peaks $\sim$~day 8, about the time of the first appearance of
the soft X-rays, and has a secondary maximum at the X-ray peak. For the forbidden lines, the [Fe\,{\sc xiv}] emission seems to track well the bright part of the SSS, fading
after the SSS drops away. [Fe\,{\sc xi}] peaks once this [Fe\,{\sc xiv}] starts to fade. Although the [Fe\,{\sc x}] emission peaks about the same time as the
SSS and the [Fe\,{\sc xiv}], it likely has multiple excitation/ionization paths. [Fe\,{\sc x}] and [Fe\,{\sc vii}] appear prior to the emergence of the SSS, and both
linger well past the fading of the [Fe\,{\sc xi}]. The uncertainties on the optical fluxes are typically $<$~5~per~cent, though up to $\sim$~10~per~cent for the faintest measurements.} 
\label{opt-lines}
\end{center}
\end{figure*}

\begin{figure}
\begin{center}
\includegraphics[clip, angle=-90, width=8.5cm]{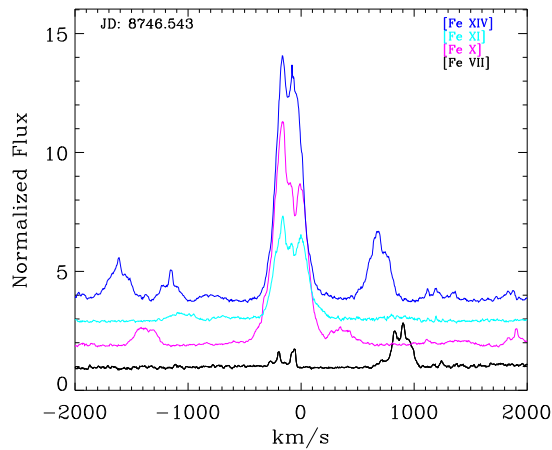}
\caption{A snapshot of the forbidden Fe lines, taken on 2019 September 20. The fluxes have been normalised to the continuum, and then offset by one continuum unit. The radial velocity of the RG is $-$90 km~s$^{-1}$.}
\label{fe}
\end{center}
\end{figure}

The fluxes of 23 discrete lines were measured in the SMARTS/Chiron spectra, and are listed in Table~\ref{smarts}. Fig.~\ref{smartsspec} shows a sample  of these spectra\footnote{See also \url{http://www.astro.sunysb.edu/fwalter/SMARTS/NovaAtlas/v3890sgr/spec/v3890sgr.chspec.html}}, while typical light-curves are shown in Fig.~\ref{opt-lines}. Based on the time variation of the fluxes, we categorise the lines as follows:

\begin{description}
\item[{\bf Lines which track the ejecta.}] These include the lines of H\,{\sc i} 
(H$\alpha$ and H$\beta$) and
He~I. These lines peak within a few days of the nova outburst, and have
a secondary peak about the time of the maximum SSS luminosity on day 15.

\item[{\bf Bowen lines and He\,{\sc ii}.}]
The He\,{\sc ii} and the $\lambda$4640 Bowen complex lines peak about day 8 (coinciding with the emergence of the SSS emission), with a
secondary peak coincident with the maximum of the SSS. 
The He\,{\sc ii} line is double-peaked, with the red peak brighter, and peaks
separated by $\sim$~170~km~s$^{-1}$. 
The Bowen complex appears to be dominated by lines of
N\,{\sc iii} and contaminated by lines of Fe\,{\sc ii} (multiplets 37 and 38).
The N\,{\sc iii} $\lambda\lambda$ 4640.6, 4641.2 lines exhibit the same
line profile as He\,{\sc ii}.
There may also be some C\,{\sc iii} $\lambda\lambda$ 4650-4652 present.

\item[{\bf Singly ionized metals.}] We examined seven prominent Fe\,{\sc ii} lines.
The lowest excitation Fe\,{\sc ii} (multiplets 41 and 49)
peak within a few days of the eruption. These lines have a second maximum
about day 8, as the SSS makes its first appearance.
Higher excitation lines in
multiplet 74 and Fe\,{\sc ii}~$\lambda$6383 peak about day 8.
All the lines show a re-brightening about days 15-25, while the SSS is at its
brightest.
In addition, Mg\,{\sc ii}~$\lambda$7786 ($\xi$~=~11.5~eV) peaks with the nova outburst;
Si\,{\sc ii}~$\lambda$6347 ($\xi$~=~8.1~eV) peaks days 7--8; and
S\,{\sc ii}~$\lambda$6084 ($\xi$~=~17.3~eV) peaks days 15--22, coincident with the SSS maximum.

\item[{\bf Forbidden Fe lines.}] 
The [Fe\,{\sc vii}] $\lambda\lambda$ 5721, 6087 lines peak at days 7--8. 
[Fe\,{\sc x}] $\lambda$6374 grows gradually until day 20, and the
coronal lines [Fe\,{\sc xi}]~$\lambda$7891 and [Fe\,{\sc xiv}]~$\lambda$5303
peak on days 18--24, following the peak of the SSS luminosity.
The [Fe\,{\sc xiv}] line peaks about 4 days before the [Fe\,{\sc xi}] line, and
the [Fe\,{\sc x}] line is the last to decay. A snapshot example of the forbidden Fe lines, shown in Fig.~\ref{fe}, demonstrates the respective strengths of the lines.

\end{description}

The first peak in the emission line fluxes (He\,{\sc i}, H\,{\sc i}, Fe\,{\sc ii}), immediately following
the eruption, is likely due directly to
the nova outburst and ejecta interacting with the donor star.
These lines are shifted by $-$90~km~s$^{-1}$, which is the radial velocity of the RG donor prior to the outburst; the ejecta themselves have a much higher
velocity. The trend of Mg\,{\sc ii}~$\lambda$7786 and $\lambda$2800 (Fig.~\ref{uvlines}) fluxes is consistent with this.

The second peak in line fluxes about day 8 coincides with the onset of the SSS. 
As the optical depth to the X-ray-emitting region decreases, permitted
optical lines
pumped by He\,{\sc ii}~$\lambda$304 become visible. This is the so-called Bowen
mechanism \citep{bowen, kall80}. These lines include O\,{\sc iii}~$\lambda\lambda$~3133, 3444
(Fig.~\ref{uvlines}), N\,{\sc iii}~$\lambda$4640, and He\,{\sc ii}~$\lambda$4686.
 He\,{\sc ii} becomes fully
ionized, decreasing the Bowen pumping and the He\,{\sc ii} fraction.

The [Fe\,{\sc xiv}] emission begins to rise about
day 10, shortly after the onset of the SSS, and peaks about day 20, some three days
after the peak SSS emission. As this recombines, the [Fe\,{\sc xi}] emission
peaks about four days later, after the SSS emission has faded by over an order
of magnitude. While the peak [Fe\,{\sc x}] emission occurs between these two
lines, it lingers longer. The profiles of the line fluxes, modelled as Gaussians, give FWHM of 5, 6, and 10 days for [Fe\,{\sc xiv}], [Fe\,{\sc xi}], and
[Fe\,{\sc x}], respectively.
This behaviour is consistent with a scenario wherein
the wind/extended atmosphere of the RG is flash-ionized by the 
brightening of the SSS. 

In the case of the $BVI$ bands, $\sim$~80--90~per~cent of the flux is in the continuum, rather than in line emission; the $R$-band flux is, however, dominated by H$\alpha$. The peak line contribution occurs around days 25--30 after eruption, interestingly about the time when the X-ray BB luminosity peaks and the temperature is coolest (Fig.~\ref{spec-fit}).

Integrating the SSS emission, and assuming a
typical 0.5~keV photon energy, the ionizing X-ray fluence is about
7~$\times$~10$^{51}$~photons. Following the discussion in Section~\ref{red}, with a solar Fe
abundance of 7.63, we estimate there are of order 10$^{44}$ Fe atoms in the
stellar atmosphere/wind, so there are sufficient X-ray photons to ionize the
wind. 
The radiative recombination time of
[Fe\,{\sc xiv}] is 0.017~s \citep{agg14}, and integrating the [Fe\,{\sc xiv}] flux requires about 7~$\times$~10$^{50}$ de-excitations
over 9--10 days. The recombination time following the turnoff of the SSS about
day 25 suggests an electron density of order 1000~cm$^{-3}$.

\section{Discussion}
\label{disc}

The data collected by {\em Swift} and {\em NICER} of the latest eruption of V3890~Sgr add to the sample of novae well monitored over the X-ray band. The frequent observations allow us to follow the X-ray emission as the source evolves rapidly from a hard, shock spectrum, through the supersoft phase, and on towards a return to quiescence. UV data from {\em Swift}, combined with optical measurements from the AAVSO, provide well-sampled light-curves, with the UVOT grism and optical spectra from SMARTS showing how the spectra change with time.

\subsection{X-ray emission}

The fitted BB temperature in Fig.~\ref{spec-fit} started off cool, rising by $\sim$~50~eV during the four days the target could not be observed by {\em Swift}. After reaching a peak of $\sim$~90--100~eV, the BB temperature followed a declining trend until the component was no longer statistically significant. The hotter optically-thin thermal component was also seen to cool, from $\go$~10~keV at the time of the earliest detection until around 20 days after eruption, after which the temperature of this component remained around 1--2~keV. When detectable, the cooler optically-thin component was consistently around 0.5--1~keV.

The BB luminosity and radius were found to increase by a factor of 10--15 as the soft emission temperature decreased, after an initial drop during the unobservable interval. From the theory of nova evolution \citep{mac85}, it is expected that the effective radius decreases (and kT increases) during the start of the SSS emission, as we see deeper into the hotter layers. In V3890~Sgr, this phase appears to have occurred rapidly, between days 9 and 13 after eruption, followed by a monotonic cooling.

\citet{page14} present a sample of \emph{Swift}-XRT spectra of nova SSS phases \citep[see also individual studies in, for example,][]{page10, julo11, page15b, bode16, aydi18b}. While many of these systems do show an initial rise in temperature as the soft emission becomes visible, if there is a later cooling detected, this is typically seen as the X-ray emission fades away. In the case of V3890~Sgr, however, the steady cooling started before the X-ray count rate had even peaked. It should be noted though, that, unlike in many nova outbursts \citep[see, e.g.,][]{page20}, there is not a clearly defined peak X-ray interval in the light-curve of V3890~Sgr, with the flux variability (which often signifies the onset of the SSS emission) transitioning immediately into the fading regime. 

The BB temperature measured for V3890~Sgr dropped very rapidly, cooling by a factor of $\sim$~3 from the peak temperature over 13 days. In comparison, the temperature of the soft component in RS~Oph decreased by only a factor of 1.5 over an interval of 25 days \citep[see][]{page14}.

Using the luminosity-temperature relations from \citet{wolf13} and \citet{sala05} with the peak SSS temperature of 90--100~eV, we estimate the WD mass in V3890~Sgr to be $\sim$~1.25--1.3$\msun$. The first indication of the soft emission being as early as day 8.5 after eruption, and the short duration of the SSS (a soft component was no longer statistically significant beyond day 26) are further signs of a massive WD \citep{kraut96,mac96,kato97,hachisu07}, as expected for a recurrent nova system.

\subsection{QPO}


Short period QPOs have been detected by {\em Swift}, {\em XMM-Newton} and {\em Chandra} \citep{ness15}, in both novae and persistent super-soft sources. In all cases, these modulations are not strictly periodic (hence the term `quasi-periodic' oscillations), have amplitudes of up to $\sim$~15--20~per~cent of the source count rate (though frequently lower), and are transient in nature (sometimes persisting for only $\sim$~120~s, but other times remaining detected for an entire {\em Swift} pointing of $\sim$~1.5~ks; Beardmore et al. in prep). As noted in \citet{andy08} and \citet{julo11}, there is no obvious correlation between the period and the source intensity. V3890~Sgr adds another source to the sample of X-ray QPOs detected in SSS novae, with a period of $\sim$~83~s and modulation amplitude of $\sim$~5~per~cent.

The cause of these QPOs is not certain, with various possibilities being suggested, from rotation to pulsation \citep[see ][for more discussion]{ness15}.
WDs in cataclysmic variables (CVs) can have rapid rotation periods of $<$~100~s -- see the Ritter \& Kolb CV catalogue\footnote{RKcat Edition 7.24 2015} \citep{ritter} -- and dwarf nova oscillations (DNOs) can also be around this duration \citep{cordova84, mason88, jones92}. However, in order to be able to detect a modulation, there needs to be some form of differentiation of the emission -- be this the occultation of a magnetic hot spot, variable absorption, or some other mechanism.
It is believed that SSS emission originates in an extended nuclear-burning atmosphere, which would then require some form of asymmetry for a modulation to be detected. Work by Beardmore et al. (in prep) does suggest that, in the case of RS~Oph, the modulation may be caused by variations in the oxygen column density, although the other novae investigated do not show such an absorption effect.
While \citet{julo11} postulated that the non-radial g-mode pulsations predicted by \cite{kawaler88} for planetary nebulae nuclei might be the cause of the QPO, recent work by \citet{wolf18} only predicts stable pulsations with periods under $\sim 10$~s, shorter than the modulations detected in SSS novae. Stellar pulsations do not, therefore, appear to be a viable 
explanation. Thus, at the present time, while QPOs are frequently detected in the SSS emission of X-ray bright novae, the formation mechanism remains unclear.

\subsection{UV and optical light-curve evolution}
\label{breaks}

As Fig.~\ref{lc} shows, while the UV and optical light-curves follow a monotonic decay (in contrast to the X-rays), the rate of fading varies over time, first steepening, then flattening after $\sim$~day 30. The decay indices can be estimated following \cite{page13}, parametrizing the evolution of the magnitudes as being proportional to log(time), equivalent to flux proportional to time; i.e. $f~\propto~(t/1~d)^{-\alpha}$. Considering only the data after day 3 (that is, the time after which the UV source had faded sufficiently to allow standard UVOT photometry), a model with four breaks follows the overall decay well; these break times, identified by eye, with uncertainties of $\lo$~0.5~day, are marked with vertical lines in the AAVSO panel of Fig.~\ref{lc}. While additional changes in slope may parametrize the decay even more exactly, these four breaks after day 3 were very obvious and, as discussed below, coincide with changes in the X-ray emission. We do note that the earliest AAVSO data follow a steeper slope than after day 3, though. 
We consider only the {\em Swift}-$uvm2$ and AAVSO-$V$ light-curves, since the $uvw2$ data are much sparser; Table~\ref{uvoptlc} lists the power-law decay slopes and break times.

\begin{table}

\caption{Parametrization of the UV and optical light-curves after day 3 as a series of power-laws. Breaks are given as days since 2019 August 27.75. The uncertainties on the break times are $\lo$~0.5~day. The statistical errors on the magnitudes are generally below 0.5~per~cent.}

\begin{center}
\begin{tabular}{lcccccccccc}
\hline
Filter & $\alpha_1$ & Break & $\alpha_2$ & Break & $\alpha_3$ & Break & $\alpha_4$ & Break & $\alpha_5$\\ 
\hline
$V$ & 0.8 & 8 & 2 & 21.5 & 5.8 & 30 & 2.8 & 40.5 & 1.4\\
$uvm2$ & 0.8 & 13 & 2 & 21.5 & 4.6 & 30 & 2.8 & 40.5 & 1.8\\
\hline
\end{tabular}

\label{uvoptlc}
\end{center}
\end{table}

The approximate times of the changes in slope are found to be consistent between the AAVSO $V$-band and UVOT $uvm2$ decay, with the exception of the first break.  Given that the AAVSO data were collected continuously throughout this interval, while there was a gap in the UVOT data, this suggests the $V$-band data may be a more reliable measurement of this break time. In addition, the earliest evidence of the SSS X-ray emission was noted on day 8.5, indicating a link between  this emergence and the change in optical slope. The second steepening, around day 21--22, coincides with the time at which the SSS X-ray emission starts to fade continuously, and a flattening in the X-ray decay is apparent around day 30 (Fig.~\ref{lc}), when the UV/optical decays also become less steep. Note that, if the time of the eruption were taken to be 0.7~day earlier (that is, corresponding to the last non-detection), the estimated slopes would be $\lo$0.1 steeper.


After re-emerging from behind the Sun, the UV source was initially approximately consistent in brightness with the data collected three months earlier, though then faded by another magnitude to m2~=~17.69~$\pm$~0.11 by 2020 May 04; the AAVSO data show measurements around $V$~$\sim$~15.5--15.8 at this time. \cite{brad10} gives the quiescent $V$-band magnitude of V3890~Sgr as 15.5, suggesting that this final flattening may be due to the source reaching the inter-eruption brightness level. The X-ray 0.3--10~keV count rate decreased by an order of magnitude during the interval of Sun constraint, from $\sim$~0.3 to $\sim$~0.03~count~s$^{-1}$, fading further to (7~$\pm$~2)~$\times$~10$^{-3}$~count~s$^{-1}$ by May 04. Assuming a distance of 4.5~kpc, this corresponds to an observed 0.3--10~keV luminosity of 4.2~$\times$~10$^{32}$~erg~s$^{-1}$. We note that an upper limit of $\sim$~10$^{31}$~erg~s$^{-1}$ was estimated from {\em XMM-Newton} data in 2010 \citep{orio20}, 20 years after the previous eruption.

\cite{strope10} categorise V3890~Sgr as a `Type S' stereotypical nova light-curve, meaning the decline is a smooth series of power-laws with no major fluctuations. Many \citep[60--90~per~cent from][]{brad10} recurrent novae fall within the P-class, showing a plateau in their optical curves around 3--6 mag below peak, which typically happens at the same time as the SSS X-ray emission \citep{hachisu00, hachisu06, brad10}. It has been speculated that this flattening may arise from the re-radiation of the soft X-rays by an accretion disc. 
There is no strong evidence for such a plateau in V3890~Sgr; in fact, the decay slope in the UV/optical actually steepens at the start of the X-ray SSS phase (though does steepen further as the supersoft emission comes to an end).

Using AAVSO data from the 1990 outburst \citep[for which the peak magnitude measured was V~$\sim$~8.1, approximately 1 mag fainter than that reported for the 2019 outburst; ][]{gonz92}, \cite{strope10} parametrize the decay of V3890~Sgr with three power-law segments. They fit the decline with a power-law slope\footnote{\cite{strope10} worked in flux space; their numbers have been divided by a factor of 2.5 to convert to slopes in a magnitude-log(time) plot as presented here} of 1.6 until day 10, then 2.4 until day 33, and finally a slope of 2, fitting data out until about 70 days after outburst. These break times can be compared favourably with our measurements of day 10 and 30 (although our data are further improved with an additional break between these, at day 21). Our best fit slope before day 8--10 is flatter, with $\alpha$~$\sim$~0.8; however, figure 4 of \cite{strope10} reveals that the earliest, slowest declining data have not been included in their analysis. Given that we find an additional break around day 21 to be a significant improvement -- we do note that \cite{strope10} have very few data points in their light-curve between days 10 and 20 -- an exact comparison cannot be performed for this middle interval.  Determining an average slope (despite it being a poor fit) for our data between days 8 and 30 gives $\alpha$~$\sim$~2.8, however, which is similar to their result of 2.4. The final slope from \cite{strope10}, covering an interval of day 33--70, of $\alpha$~$\sim$~2 is within our fitted range (again with an additional break) of 2.8 to 1.4. These results imply that the recurrent explosions for a given nova are similar, but not identical.

\cite{hk06} previously proposed a universal decline law for the optical and infrared data of {\em classical} novae which produces a smooth decay of the same form as the S-class curves; \cite{strope10} suggest these slopes should also apply to RNe, though possibly with a plateau phase in the middle. The underlying template from \cite{hk06} has a slope steepening from $\alpha$~$\sim$ 1.75 to $\sim$~3.5 around 6 mag below peak, with the break explained as being due to a sudden decrease in the wind mass-loss rate (where the eruption is being modelled as a radiatively-driven wind). The decline then flattens to $\alpha$~$\sim$~3 as the wind ceases. In comparison, considering their complete sample of S-class curves, \cite{strope10} find mean slopes of 1.6 and 2.1 at early and late times, with the late-time slope being substantially flatter than that predicted by \cite{hk06}, while our fit to V3890~Sgr here is even flatter still. \cite{strope10} suggest these differences may be caused by the rise of emission lines on top of the continuum, a speculation we also made when comparing $V$-band data with other filters for V959~Mon \citep{page13}. It is likely that this proposed universal decline law is an over-simplification, however, with substantial, unexplained scatter in the slopes measured from actual data \citep{strope10}.

Links between changes in the shape of the UV/optical and X-ray light-curves have been seen in other novae as well. For example, a steepening in the UV/optical decay at the same time as the start of the fading of the super-soft emission was identified in V745~Sco \citep[another symbiotic-like RN; ][]{page15b}, and both V959 Mon \citep{page13} and V2491 Cyg \citep{page10} show breaks in the UV/optical decay rate at (close to) the same times as changes in the X-ray behaviour. These temporal coincidences are indicative of a link between the emission regions, despite the X-ray and UV/optical light-curves being distinctly different in shape before the super-soft emission has ceased.

\section{Summary}

V3890~Sgr is a recurrent nova now seen in eruption three times between 1962 and 2019.  The main parameters derived or adopted for the system for the latest eruption are summarised in Table~\ref{params}. 

{\em Swift} observations started very rapidly after the discovery of the eruption. The X-ray emission was initially found to be hard, consistent with shocks, and absorbed by a gradually decreasing column. At later times, N$_{\rm H}$ in constant excess of the Galactic value is required; this can be modelled as further absorption by the RG wind. After a slow brightening, a new soft component was detected on $\sim$~day~8.5, leading to the count rate increasing by about an order of magnitude over the following week. This SSS phase lasted fewer than 20~days which, together with the early onset and high peak temperature of the soft emission, implies a massive WD in this recurrent nova system.  While V3890~Sgr had been previously detected in the X-ray band by {\em ROSAT} \citep{orio01}, those pointings occurred many months after the 1990 eruption, so the data presented here are the first observations of the SSS phase in this recurrent nova.

A QPO of $\sim$~83~s was briefly detected during the first {\em NICER} observation, around day~10. V3890~Sgr therefore adds to the sample of super-soft novae showing quasi-periodic variations on a timescale of $<$~100~s.

Each UV grism spectrum shows a continuum superimposed by emission lines, including N, C, Mg and O; an SMC extinction law best dereddens the data. The UV/optical light-curve follows a monotonic decay, with changes in slope occurring at times where the X-ray light-curve also varies in shape. As for other novae, this suggests a link between the UV/optical and X-ray emitting regions, despite the obvious differences in the overall light-curve shape.

The flux evolution of the optical lines shows an initial peak around the time of the nova eruption, followed by a second peak coincident in time with the onset of the SSS phase. These are likely caused by the nova ejecta interacting with the RG donor, and the extended atmosphere of the RG being flash-ionized by the super-soft photons, respectively. The peak line contribution to the flux occurs around the same time that the BB is coolest and its luminosity reaches a peak.

In order to understand V3890~Sgr better, information about the inclination of the system and observations during different orbital phases, to determine whether the X-ray absorbing column changes appreciably, would be useful.

\begin{table}

\caption{Summary of parameters for V3890~Sgr.}

\begin{center}
\begin{tabular}{lcc}
\hline
Property & Value\\
\hline
2019 eruption time & 2019 August 27.75 \\
Previous eruptions & 1962, 1990\\
Recurrence time & 28--29 yr\\
Adopted distance & 4.5 kpc\\
N$_{\rm H}$ from ISM & 1.9~$\times$~10$^{21}$~cm$^{-2}$\\
Late-time excess N$_{\rm H}$ from X-ray fits & 5.1~$\times$~10$^{21}$~cm$^{-2}$\\
Adopted A$_{\rm V}$ & 1.77 \\
M$_{\rm WD}$ & $\sim$~1.3$\msun$\\
SSS emission phase & day 8.5--26\\
QPO & 83~s\\

\hline
\end{tabular}

\label{params}
\end{center}
\end{table}

\section{Data availability}

The data underlying this article are available in the archives at \url{https://www.swift.ac.uk/swift\_live/} ({\em Swift}) and \url{https://heasarc.gsfc.nasa.gov/cgi-bin/W3Browse} ({\em Swift} and {\em NICER}). XRT products can be extracted automatically at \url{https://www.swift.ac.uk/user_objects/}.
SMARTS data are included in the `Stony Brook/SMARTS Atlas of (mostly) Southern Novae' at \url{http://www.astro.sunysb.edu/fwalter/SMARTS/NovaAtlas/atlas.html}. AAVSO data are available from \url{https://www.aavso.org/data-download}.

\section*{ACKNOWLEDGEMENTS}
\label{ack}

KLP, APB, NPMK and JPO acknowledge funding from the UK Space Agency. KVS acknowledges support from NSF award AST-1751874, NASA award 11-Fermi 80NSSC18K1746 and a Cottrell fellowship of the Research Corporation. FMW acknowledges support of the US taxpayers through NSF grant 1611443. Access to the SMARTS
partnership is made possible in part by research support from Stony Brook University.
This work made use of data supplied by the UK Swift Science Data Centre at the University of Leicester. We thank the following observers for contributing their V-band data to the AAVSO: SKA, MRV, FRF, OCN, JPG, SFLB, TUB, LGIB, HMB, SHS, CDAD, FJAA, KNAA, BHQ, MGW, VMT, RZD, GPX, ATE, BSEC, LJEC, SPET, RBRC, MBRB, NLX.

\appendix

\section{Photometry}

Table~\ref{photom} lists the {\em Swift}-UVOT photometry used in this paper. The AAVSO-optical data points were taken directly from \url{https://www.aavso.org/data-download}. Table~\ref{smartslog} provides the SMARTS observing log, taken from \url{http://www.astro.sunysb.edu/fwalter/SMARTS/NovaAtlas/v3890sgr/v3890sgr.html}.

\begin{table}

\caption{UV photometry obtained by {\em Swift}-UVOT. The times are given in days since 2019 August 27.75. }

\begin{center}
\begin{tabular}{lcc}
\hline
Filter & Day & Magnitude\\
\hline
$u$ & 1.33 & 8.84~$\pm$~0.04\\
$u$ &  2.20 & 9.19~$\pm$~0.04\\
\hline
$uvw1$ & 0.66 & 8.59~$\pm$~0.02\\
$uvw1$ & 0.72 & 8.62~$\pm$~0.02\\
\hline
$uvm2$ & 3.2534 & 11.05~$\pm$~0.02\\
$uvm2$ & 4.0529 &	11.38~$\pm$~0.02\\
$uvm2$ & 4.3855 &	11.47~$\pm$~0.02\\
$uvm2$ & 4.6387 &	11.53~$\pm$~0.02\\
$uvm2$ & 5.9827 &	11.77~$\pm$~0.02\\
$uvm2$ & 6.7060 &	11.82~$\pm$~0.02\\
$uvm2$ & 6.8375 &	11.87~$\pm$~0.02\\
$uvm2$ & 7.5766 &	11.88~$\pm$~0.02\\
$uvm2$ & 7.7755 &	11.90~$\pm$~0.02\\
$uvm2$ & 13.2810 &	12.46~$\pm$~0.02\\	
$uvm2$ & 13.4729 &	12.52~$\pm$~0.03\\	
$uvm2$ & 13.7385 &	12.54~$\pm$~0.03\\	
$uvm2$ & 15.2176 &	12.70~$\pm$~0.02\\	
$uvm2$ & 15.3495 &	12.69~$\pm$~0.02\\	
$uvm2$ & 15.5491 &	12.69~$\pm$~0.02\\	
$uvm2$ & 15.8014 &	12.75~$\pm$~0.02\\	
$uvm2$ & 16.1471 &	12.78~$\pm$~0.02\\	
$uvm2$ & 16.2660 &	12.83~$\pm$~0.02\\	
$uvm2$ & 16.8619 &	12.83~$\pm$~0.02\\	
$uvm2$ & 17.0617 &	12.86~$\pm$~0.02\\	
$uvm2$ & 17.7412 &	12.92~$\pm$~0.02\\	
$uvm2$ & 17.8680 &	12.94~$\pm$~0.02\\     
$uvm2$ & 18.1398 &	12.97~$\pm$~0.02\\
$uvm2$ & 18.3908 &	12.96~$\pm$~0.02 \\    
$uvm2$ & 18.8550 &	13.00~$\pm$~0.03\\
$uvm2$ & 19.0603 &	13.02~$\pm$~0.02\\	
$uvm2$ & 19.2614 &        13.05~$\pm$~0.02\\
$uvm2$ & 19.7274 &	13.12~$\pm$~0.02\\	
$uvm2$ & 19.9842 &	13.13~$\pm$~0.03\\	
$uvm2$ & 20.1321 &	13.12~$\pm$~0.02\\	
$uvm2$ & 23.5044 &	13.63~$\pm$~0.03\\	
$uvm2$ & 23.5114 &	13.66~$\pm$~0.03\\	
$uvm2$ & 23.8542 &	13.73~$\pm$~0.03\\	
$uvm2$ & 24.0438 &	13.74~$\pm$~0.03\\	
$uvm2$ & 24.1187 &	13.75~$\pm$~0.03\\	
$uvm2$ & 24.0483 &	13.73~$\pm$~0.03\\	
$uvm2$ & 24.3751 &	13.77~$\pm$~0.03\\	
$uvm2$ & 24.7644 &	13.79~$\pm$~0.08\\	
$uvm2$ & 25.2325 &	13.98~$\pm$~0.03\\    
$uvm2$ & 25.4284 &	14.04~$\pm$~0.03\\
$uvm2$ & 25.8942 &	14.11~$\pm$~0.03\\	
$uvm2$ & 25.9609 &	14.15~$\pm$~0.03\\	
$uvm2$ & 26.2962 &	14.19~$\pm$~0.03\\	
$uvm2$ & 26.6264 &	14.27~$\pm$~0.03\\	
$uvm2$ & 27.2236 &	14.39~$\pm$~0.03\\	
$uvm2$ & 27.5573 &	14.46~$\pm$~0.03\\	
$uvm2$ & 27.6907 &	14.47~$\pm$~0.03\\	
$uvm2$ & 27.9555 &	14.53~$\pm$~0.03\\
$uvm2$ & 28.5570 &	14.64~$\pm$~0.03\\	
$uvm2$ & 28.6913 &	14.64~$\pm$~0.03\\	
$uvm2$ & 28.9502 &	14.68~$\pm$~0.03\\
$uvm2$ & 29.5613 &	14.81~$\pm$~0.03\\	
$uvm2$ & 29.9566 &	14.89~$\pm$~0.04\\	
$uvm2$ & 30.8882 &	15.00~$\pm$~0.03\\
$uvm2$ & 32.2015 &	15.14~$\pm$~0.03\\
$uvm2$ & 33.1290 &	15.22~$\pm$~0.04\\
$uvm2$ & 33.8051 &	15.29~$\pm$~0.03\\
$uvm2$ & 35.7278 &	15.52~$\pm$~0.05\\
$uvm2$ & 36.5298 &	15.49~$\pm$~0.03\\
\end{tabular}
\end{center}
\end{table}
\addtocounter{table}{-1}
\begin{table}
\begin{center}
\caption{-- {\bf continued}} 
\begin{tabular}{lcc}
\hline
Filter & Day & Magnitude\\
\hline
$uvm2$ & 43.2252 &	16.04~$\pm$~0.04\\
$uvm2$ & 44.3510 &	16.03~$\pm$~0.06 \\
$uvm2$ & 44.7490 &	16.19~$\pm$~0.07\\	
$uvm2$ & 45.3512 &	16.04~$\pm$~0.07\\	
$uvm2$ & 46.6153 &	16.15~$\pm$~0.05\\
$uvm2$ & 48.5360 &	16.21~$\pm$~0.04\\
$uvm2$ & 51.5312 &	16.39~$\pm$~0.04\\
$uvm2$ & 55.1820 &	16.50~$\pm$~0.05\\
$uvm2$ & 61.6879 &	16.75~$\pm$~0.05\\
$uvm2$ & 68.7261 &	17.01~$\pm$~0.06\\	
$uvm2$ & 75.2993 &	17.11~$\pm$~0.06\\
$uvm2$ & 173.3463 &	16.69~$\pm$~0.04\\
$uvm2$ & 177.1258 &	16.96~$\pm$~0.05\\
$uvm2$ & 184.4007 &	16.94~$\pm$~0.04\\
$uvm2$ & 196.6866 &	17.39~$\pm$~0.05\\
$uvm2$ & 199.3379 &	17.28~$\pm$~0.05\\
$uvm2$ & 245.2738 &	17.72~$\pm$~0.10\\
$uvm2$ & 250.0637 &	17.69~$\pm$~0.11\\
\hline
$uvw2$ & 3.2466 &	10.97~$\pm$~0.02\\
$uvw2$ & 5.3267 & 	11.11~$\pm$~0.04\\
$uvw2$ & 5.9906 &	11.25~$\pm$~0.03\\
$uvw2$ & 6.8461 &	11.31~$\pm$~0.03\\
$uvw2$ & 7.5842 &	11.35~$\pm$~0.03\\
$uvw2$ & 7.7835 &	11.34~$\pm$~0.03\\
$uvw2$ & 8.5144 &	11.35~$\pm$~0.03\\
$uvw2$ & 9.2434 &	11.32~$\pm$~0.02\\
$uvw2$ & 16.5343 &	12.02~$\pm$~0.03\\
$uvw2$ & 16.5406 &	11.93~$\pm$~0.04\\
$uvw2$ & 17.4717 &	12.06~$\pm$~0.03\\
$uvw2$ & 17.4774 &	12.00~$\pm$~0.05\\
$uvw2$ & 18.5194 &	12.21~$\pm$~0.02\\
$uvw2$ & 18.5253 &	12.12~$\pm$~0.04\\
$uvw2$ & 18.5309 &	12.23~$\pm$~0.04\\
$uvw2$ & 19.9912 &	12.35~$\pm$~0.03\\
$uvw2$ & 19.9968 &	12.36~$\pm$~0.04\\
$uvw2$ & 21.1828 &	12.52~$\pm$~0.03\\
$uvw2$ & 21.1891 &	12.54~$\pm$~0.04\\
$uvw2$ & 21.6469 &	12.57~$\pm$~0.05\\
$uvw2$ & 21.7795 &	12.59~$\pm$~0.04\\
$uvw2$ & 22.9802 &	12.85~$\pm$~0.02\\
$uvw2$ & 22.9876 &	12.82~$\pm$~0.03\\
$uvw2$ & 23.0462 &	12.87~$\pm$~0.02\\
$uvw2$ & 23.0536 &	12.85~$\pm$~0.03\\
$uvw2$ & 24.0558 &	13.20~$\pm$~0.05\\
$uvw2$ & 25.1148 &	13.44~$\pm$~0.05\\
$uvw2$ & 25.1697 &	13.42~$\pm$~0.05\\
\hline
\end{tabular}

\label{photom}
\end{center}
\end{table}

\begin{table}

\caption{Complete SMARTS spectroscopic observing log for 2019. All spectra were obtained using the Chiron spectrograph. The early `slicer mode' observations were continuum-dominated, and not relevant to this investigation.}

\begin{center}
\begin{tabular}{lccccc}
\hline
Date & JD & Mode & Resolution &   Wavelength & Exposure  \\
(UT) & (245+)& & & range (\AA)  & time (s)\\
\hline
  2019 Aug. 28 & 8723.526 & slicer &  78000 &  4071--8770 &    600  \\   
  2019 Aug. 29 & 8724.556 & slicer &  78000 &  4071--8770 &    900 \\    
  2019 Aug. 30 & 8725.576 & slicer &  78000 &  4071--8770 &   1200  \\   
  2019 Aug. 1 & 8726.518 & slicer &  78000 &  4071--8770 &   1200  \\   
  2019 Sep. 01 & 8727.553 & slicer &  78000 &  4071--8770 &   1200  \\   
  2019 Sep. 02 & 8728.588 & fiber &   27800 &  4070--8907 &    600  \\   
  2019 Sep. 03 & 8729.542 & fiber &   27800 &  4070--8907 &    600 \\    
  2019 Sep. 04 & 8730.531 & fiber &   27800 &  4070--8907  &   600 \\    
  2019 Sep. 05 & 8731.516 & fiber &   27800 &  4070--8907 &    600 \\    
  2019 Sep. 08 & 8734.561 & fiber &   27800 &  4070--8907 &    600 \\    
  2019 Sep. 12 & 8738.528 & fiber &   27800 &  4070--8907  &   900  \\   
  2019 Sep. 13 & 8739.492 & fiber &   27800 &  4070--8907  &   900 \\    
  2019 Sep. 14 & 8740.497 & fiber &   27800 &  4070--8907  &   900 \\    
  2019 Sep. 15 & 8741.511 & fiber &   27800 &  4070--8907  &   900 \\    
  2019 Sep. 16 & 8742.521 & fiber &   27800 &  4070--8907  &   900 \\    
  2019 Sep. 17 & 8743.483 & fiber &   27800 &  4070--8907 &   1200 \\    
  2019 Sep. 18 & 8744.517 & fiber &   27800 &  4070--8907 &   1200  \\   
  2019 Sep. 19 & 8745.519 & fiber &   27800 &  4070--8907 &   1200 \\    
  2019 Sep. 20 & 8746.543 & fiber &   27800 &  4070--8907 &   1200 \\    
  2019 Sep. 21 & 8747.628 & fiber &   27800 &  4070--8907 &   1500 \\    
  2019 Sep. 22 & 8748.582 & fiber &   27800 &  4070--8907 &   1500 \\    
  2019 Sep. 24 & 8750.576 & fiber &   27800 &  4070--8907 &   1500 \\    
  2019 Sep. 25 & 8751.511 & fiber &   27800 &  4070--8907 &   1500 \\    
  2019 Sep. 26 & 8752.515 & fiber &   27800 &  4070--8907 &   1800 \\    
  2019 Oct. 02 & 8758.543 & fiber &   27800 &  4070--8906 &   2700 \\    
  2019 Oct. 05 & 8761.530 & fiber &   27800 &  4070--8906  &  2700 \\    
  2019 Oct. 08 & 8764.500 & fiber &   27800 &  4070--8906  &  2700 \\    
  2019 Oct. 11 & 8768.479 & fiber &   27800 &  4070--8906  &  2700 \\    
  2019 Oct. 15 & 8772.483 & fiber &   27800 &  4070--8907 &   2700 \\    
  2019 Oct. 20 & 8776.508 & fiber &   27800 &  4070--8907 &   2700 \\    
  2019 Oct. 27 & 8784.491 & fiber &   27800 &  4070--8907 &   3600 \\    
  2019 Nov. 04 & 8792.496 & fiber &   27800 &  4070--8907 &   3600 \\    

\hline
\end{tabular}

\label{smartslog}
\end{center}
\end{table}

\bsp	
\label{lastpage}

\begin{thebibliography}{}

\bibitem[Aggarwal \& Keenan (2014)]{agg14}Aggarwal K.M., Keenan F.P., 2014, MNRAS, 445, 2015
\bibitem[Arnaud (1996)]{arn96}Arnaud K. A., 1996, in Jacoby G., Barnes J., eds, ASP Conf. Ser. Vol. 101, Astronomical Data Analysis Software and Systems V. Astron. Soc. Pac., San Francisco, p. 17
\bibitem[Aydi et al. (2018a)]{aydi18a}Aydi E. et al., 2018a, MNRAS, 474, 2679
\bibitem[Aydi et al. (2018b)]{aydi18b}Aydi E. et al., 2018b, MNRAS, 480, 572
  
\bibitem[Beardmore et al. (2019a)]{beardmore19a}Beardmore A.P., Page K.L., Markwardt C.B., Gendreau K.C., Arzoumanian Z., Pope J.S., 2019a, Astron. Telegram, 13086
\bibitem[Beardmore et al. (2019b)]{beardmore19b}Beardmore A.P., Osborne J.P., Page K.L., Ness J.U., Orio M., Drake J.J., 2019b, Astron. Telegram, 13104
\bibitem[Beardmore, Osborne \& Page (2013)]{andy13}Beardmore A.P. Osborne A.P., Page K.L., 2013, Astron. Telegram, 5573
\bibitem[Beardmore et al. (2010)]{andy10}Beardmore A.P. et al., 2010, Astron. Telegram, 2423
\bibitem[Beardmore et al. (2008)]{andy08}Beardmore A.P., Osborne J.P., Page K.L., Goad M.R., Bode M.F., Starrfield, S., 2008, in  Evans A., Bode M. F., O’Brien T.J., Darnley M. J., eds, ASP Conf. Ser. Vol. 401, RS Ophiuchi (2006) and the Recurrent Nova Phenomenon. Astron. Soc. Pac., San Francisco, p. 296
\bibitem[Bode \& Evans (2008)]{bode08}Bode M.F., Evans A., 2008, Classical Novae, 2nd edn., Cambridge Astrophysics Series 43, Cambridge Univ. Press, Cambridge
\bibitem[Bode \& Kahn (1985)]{bode85}Bode, M.F., Kahn F.D., 1985, MNRAS, 217, 205
\bibitem[Bode et al. (2006)]{bode06}Bode M.F. et al. 2006, ApJ, 652, 629
\bibitem[Bode et al. (2016)]{bode16}Bode M.F. et al. 2016, ApJ, 818, 145
\bibitem[B{\"o}hm-Vitense (1981)]{bv81}B{\"o}hm-Vitense E., 1981, ARA\&A, 19, 295
\bibitem[Bowen (1934)]{bowen}Bowen I.S, 1934, PASP, 46, 146
\bibitem[Burrows et al. (2005)]{bur05}Burrows D.N. et al., 2005, Space Sci. Rev., 120, 165
\bibitem[Buson, Jean \& Cheung (2019)]{buson19}Buson S., Jean P., Cheung C.C., 2019, Astron. Telegram, 13114
\bibitem[Cardelli, Clayton \& Mathis (1989)]{ccm89}Cardelli J.A., Clayton G.C., Mathis J.S., 1989, ApJ, 345, 245
\bibitem[Cash (1979)]{cash79}Cash W., 1979, ApJ, 228, 939
\bibitem[C{\'o}rdova et al. (1984)]{cordova84}C{\'o}rdova F.A., Chester T.J., Mason K.O., Kahn S.M., Garmire G.P., 1094, ApJ, 278, 739
\bibitem[Darnley et al. (2014)]{darn14}Darnley M.J., Williams S.C., Bode M.F., Henze M., Ness J.-U., Shafter A.W., Hornoch K., Votruba V., 2014, A\&A, 563, L9
\bibitem[Darnley et al. (2012)]{darn12}Darnley M.J., Ribeiro V.A.R.M., Bode M.F., Hounsell R.A., Williams R.P., 2012, ApJ, 746, 61
\bibitem[den Herder et al. (2001)]{rgs}den Herder J.W. et al., 2001, A\&A, 365, L7
\bibitem[Dinerstein \& Hoffleit (1962)]{1962}Dinerstein H., Hoffleit D., 1962, IBVS, 845, 1
\bibitem[Evans et al. (2019)]{evans19}Evans A., Banerjee D.P.K., Geballe T.R., Joshi V., Woodward C.E., Gehrz R.D., 2019, Astron. Telegram, 13088
\bibitem[Evans et al. (2009)]{evans09}Evans P.A. et al., 2009, MNRAS, 397, 1177
\bibitem[Evans et al. (2007)]{evans07}Evans P.A. et al., 2007, A\&A, 469, 379
\bibitem[Fitzpatrick (1999)]{fitz99}Fitzpatrick E.L., 1999, PASP, 111, 63
\bibitem[Gebbie \& Thomas (1968)]{geb68}Gebbie K.B., Thomas R.N., 1968, ApJ, 154, 285
\bibitem[Gehrels et al. (2004)]{geh04}Gehrels N. et al., 2004, ApJ, 611, 1005 
\bibitem[Gendreau, Arzoumanian \& Okajima (2012)]{nicer}Gendreau K.C., Arzoumanian Z., Okajima, T., 2012, SPIE, 8443, 13
  \bibitem[Gendreau et al. (2016)]{nicer16}Gendreau K.C. et al., 2016, SPIE, 9905, 1
\bibitem[Gonzalez-Riestra (1992)]{gonz92}Gonzalez-Riestra R., 1992, A\&A, 265, 71
\bibitem[Green et al. (2019)]{green19}Green G.M., Schlafly E.F., Zucker C., Speagle J.S., Finkbeiner D.P., 2019, ApJ, 887, 93
\bibitem[G{\" u}ver \& {\" O}zel (2009)]{go09}G{\" u}ver T., {\" O}zel F., 2009, MNRAS, 400, 2050
\bibitem[Hachisu \& Kato (2006)]{hk06}Hachisu I. Kato M., 2006, ApJS, 167, 59
\bibitem[Hachisu, Kato \& Luna (2007)]{hachisu07}Hachisu I., Kato M., Luna G.J.M., 2007, ApJ, 659, L153
\bibitem[Hachisu et al. (2000)]{hachisu00}Hachisu I., Kato M., Kato T., Matsumoto, K., 2000, ApJ, 528, L97
\bibitem[Hachisu et al. (2006)]{hachisu06}Hachisu I et al., 2006, ApJ, 651, L141
\bibitem[Hagen et al. (2017)]{lea17}Hagen L.M.Z., Siegel M.H., Hoversten E.A., Gronwall C., Immler, S., Hagen A., 2017, MNRAS, 466, 4540
\bibitem[Harkness (1983)]{harkness83}Harkness R.P., 1983, MNRAS, 204, 45
\bibitem[Harrison, Johnson \&  Spyromilio (1993)]{harr93}Harrison T.E., Johnson J.J.,  Spyromilio J., 1993, AJ, 105, 320
 \bibitem[Hauschildt et al. (1992)]{hauschildt92} Hauschildt P.H., Wehrse R., Starrfield S., Shaviv G., 1992, ApJ, 393, 307
\bibitem[Henze et al. (2011)]{hen11}Henze M. et al., 2011, A\&A, 533, A52
\bibitem[Jones \& Watson (1992)]{jones92}Jones M.H., Watson M.G., 1992, MNRAS, 257, 633
\bibitem[Kalberla \& Haud (2015)]{kal15} Kalberla P.M.W., Haud U., 2015, A\&A, 578, A78
\bibitem[Kalberla et al. (2005)]{kal05}Kalberla P.M.W., Burton W.B., Hartmann D., Arnal E.M., Bajaja E., Morras R., P{\" o}ppel W.G.L., 2005, A\&A, 440, 775
\bibitem[Kallman \& McCray (1980)]{kall80}Kallman T., McCray R., 1980, ApJ, 242, 615
\bibitem[Kato (1997)]{kato97}Kato M., 1997, ApJS, 113, 121
\bibitem[Kawaler (1988)]{kawaler88}Kawaler S.D., 1998, ApJ, 334, 220

\bibitem[Kilmartin et al. (1990)]{1990}Kilmartin P., Gilmore A., Jones A.F., Pearce A., 1990, IAU Circ., 5002, 1
\bibitem[Krautter (2008)]{kraut08}Krautter J., 2008, in `Classical Novae', 2nd edn., Eds. M.F. Bode \& A. Evans, Cambridge Astrophysics Series 43, Cambridge Univ. Press, Cambridge, p.232
\bibitem[Krautter et al. (1996)]{kraut96}Krautter J., {\" O}gelman H,  Starrfield S., Wichmann R., Pfeffermann E., 1996, ApJ, 456, 788
\bibitem[Kuin et al. (2015)]{paul15}Kuin N.P.M. et al., 2015, MNRAS, 449, 2514
\bibitem[Kuin (2014)]{paul14}Kuin N.P.M., 2014, Astrophysics Source Code  Library, record ascl:1410.004
\bibitem[Kuin et al. (2019)]{kuin19}Kuin P. et al., 2019, Astron. Telegram, 13072
\bibitem[Lallement et al. (2014)]{lal14}Lallement R., Vergely J.-L., Valette B., Puspitarini L., Eyer L., Casagrande L., 2014, A\&A, 561, A91
  \bibitem[Leahy et al. (1987)]{leahy87} Leahy D.A., 1987, A\&A, 180, 275
\bibitem[Leahy et al. (1983)]{leahy83} Leahy D.A., Darbro W., Elsner R.F., Weisskopf M.C., Sutherland P.G., Kahn S., Grindlay J.E., 1983, ApJ, 266, 160
\bibitem[MacDonald (1996)]{mac96}MacDonald J., 1996, in Cataclysmic Variable and Related Objects, ed. A. Evans \& J.H. Wood (Dordrecht: Kluwer), ASSL, 208, 281
\bibitem[MacDonald (1985)]{mac85}MacDonald J., Fujimoto M.Y., Truran J.W., 1985, ApJ, 294, 263
\bibitem[Mason et al. (1988)]{mason88}Mason K.O., C{\`o}rdova F.A., Watson M.G., King A.R., 1988, MNRAS, 232, 779  
 \bibitem[Mr{\' o}z et al. (2014)]{mroz14}Mr{\' o}z P. et al., 2014, MNRAS, 443, 784
  \bibitem[Munari \& Walter (2019a)]{mun19}Munari U., Walter, F.M., 2019a, Astron. Telegram, 13069
\bibitem[Munari \& Walter (2019b)]{mun19b}Munari U., Walter, F.M., 2019a, Astron. Telegram, 13081
\bibitem[Ness (2020)]{ness19a}Ness J.-U., 2020, AdSpR, 66, 1202 
\bibitem[Ness et al. (2019)]{ness19b}Ness J.-U. et al., 2019, Astron. Telegram, 13124
\bibitem[Ness et al. (2015)]{ness15}Ness J.-U. et al., 2015, A\&A, 578, A39
\bibitem[Ness et al. (2013)]{ness13}Ness J.-U. et al., 2013, A\&A, 559, A50

\bibitem[Ness et al. (2009)]{ness09}Ness J.-U. et al., 2009, ApJ, 137, 4160

\bibitem[Nyamai et al. (2019)]{nya19}Nyamai M.M., Woudt P.A., Ribeiro V.A.R.M., Chomiuk L., 2019, Astron. Telegram, 13089
\bibitem[Orio et al. (2020)]{orio20}Orio M. et al., 2020, ApJ, 895, 80
\bibitem[Orio et al. (2019)]{orio19}Orio M. et al., 2019, Astron. Telegram, 13083
\bibitem[Orio, Covington \& {\" O}gelman (2001)]{orio01}Orio M., Covington J., {\" O}gelman H., 2001, A\&A, 373, 542, 
\bibitem[Osborne (2015)]{julo15}Osborne J.P., 2015, JHEAp, 7, 117
\bibitem[Osborne et al. (2011)]{julo11}Osborne J.P. et al., 2011, ApJ, 727, 124
\bibitem[Osborne et al. (2006)]{julo06}Osborne J.P. et al., 2006, Astron. Telegram, 770
\bibitem[Page \& Osborne (2014)]{page14}Page K.L., Osborne J.P., 2014, in `Stella Novae: Past and Future Decades', Eds. Woudt P.A., Ribeiro V.A.R.M., Astronomical Society of the Pacific Conference Series, Vol. 490,  345
\bibitem[Page, Beardmore \& Osborne (2020)]{page20}Page K.L., Beardmore A.P., Osborne J.P., 2020, AdSpR, 66, 1169
\bibitem[Page et al. (2019a)]{page19a}Page K.L., Beardmore A.P., Osborne J.P., Orio M., Sokolovsky K.V., Darnley M.J., 20191, Astron. Telegram, 13084
\bibitem[Page et al. (2019b)]{page19b}Page K.L., Beardmore A.P., Osborne J.P., Kuin N.P.M., Ness J.U., Orio M., Sokolovsky K.V., Starrfield S., 2019b, Astron. Telegram, 13137
\bibitem[Page et al. (2015a)]{page15a}Page K.L., Beardmore A.P., Osborne, J.P., 2015a, Astron. Telegram, 8133
\bibitem[Page et al. (2015b)]{page15b}Page K.L. et al., 2015b, MNRAS, 454, 3108
\bibitem[Page et al. (2013a)]{page13}Page K.L., Osborne J.P., Wagner R.M., Beardmore A.P., Shore S.N., Starrfield S., Woodward C.E., 2013a, ApJ, 768, L26
\bibitem[Page et al. (2010)]{page10}Page K.L. et al., 2010, MNRAS, 401, 121
\bibitem[Page et al. (2013)]{mat13}Page M.J. et al., 2013b, MNRAS, 436, 1684
\bibitem[Pei (1992)]{pei92}Pei Y.C., 1992, ApJ, 395, 130
\bibitem[Pereira (2019)]{per19}Pereira A., 2019, vsnet-alert 23505
\bibitem[Pickles (1998)]{pickles98}Pickles A.J., 1998, PASP, 110, 863
\bibitem[Polisensky et al. (2019)]{pol19}Polisensky E. et al., 2019, Astron. Telegram, 13185
\bibitem[Predehl \& Schmitt (1995)]{pre95}Predehl P., Schmitt J.H.M.M., 1995, A\&A, 293, 889
\bibitem[Queiroz et al. (2018)]{queiroz18}Queiroz A.B.A. et al., 2018, MNRAS, 476, 2556
\bibitem[Reimers (1977)]{reimers77}Reimers D., 1977, A\&A, 61, 217
\bibitem[Ritter \& Kolb (2003)]{ritter}Ritter H., Kolb U., 2003, A\&A, 404, 301
\bibitem[Robinson, Clayton \& Schaefer (2006)]{rob06}Robinson P.B., Clayton G.C., Schaefer B.E., 2006, PASP, 118, 385
\bibitem[Romano et al. (2006)]{romano06}Romano P. et al., 2006, A\&A, 456, 917
\bibitem[Roming et al. (2005)]{rom05}Roming P.W.A. et al., 2005, Space Sci. Rev., 120, 95
\bibitem[Sala \& Hernanz (2005)]{sala05}Sala G., Hernanz M., 2005, A\&A, 439, 1061
\bibitem[Schaefer (2018)]{brad18}Schaefer B.E., 2018, MNRAS, 481, 3033
\bibitem[Schaefer (2010)]{brad10}Schaefer B.E., 2010, ApJS, 187, 275
\bibitem[Schaefer (2009)]{brad09}Schaefer B.E., 2009, ApJ, 697, 721
\bibitem[Schwarz et al. (2011)]{schwarz11}Schwarz G.J. et al., 2011, ApJS, 197, 31
\bibitem[Schlafly \& Finkbeiner (2011)]{sch11} Schlafly E.F., Finkbeiner D.P., 2011, ApJ, 737, 103
\bibitem[Schlegel, Finkbeiner \& Davis (1998)]{sch98} Schlegel D,J., Finkbeiner D.P., Davis M., 1998, ApJ, 500, 525
\bibitem[Schr{\" o}der \& Cuntz (2005)]{sch05}Schr{\" o}der K.-P., Cuntz M., 2005, ApJ, 630, L73
\bibitem[Shore et al. (2011)]{steve11}Shore S.N. et al., 2011, A\&A, 527, A98
 \bibitem[Shore et al. (2012)]{steve12}Shore S.N., Wahlgren G.M., Augusteijn T., Liimets T., Koubsky P., \v{S}lechta M., Votruba, V., 2012, A\&A, 540, A55
 
\bibitem[Singh et al. (2019a)]{singh19a}Singh K.P., Girish V., Anupama G.C., Pavana M., 2019a, Astron. Telegram, 13102
\bibitem[Singh et al. (2019b)]{singh19b}Singh K.P., Girish V., Anupama G.C., Pavana M., 2019b, Astron. Telegram, 13145

\bibitem[Smith et al. (2001)]{smith01}Smith R.K., Brickhouse N.S., Liedahl D.A., Raymond J.C., 2001, ApJ, 556, L91
\bibitem[Sokolovsky et al. (2019)]{sok19}Sokolovsky K.V. et al., 2019, Astron. Telegram, 13050
\bibitem[Strader et al. (2019)]{strader19}Strader J. et al., 2019, Astron. Telegram, 13047
\bibitem[Strope, Schaefer \& Henden (2010)]{strope10}Strope R.J., Schaefer B.E., Henden A.A., 2010, AJ, 140, 34
\bibitem[Tokovinin et al. (2012)]{chiron}Tokovinin A., Fischer D.A., Bonati M., Giguere M.J., Moore P., Schwab C,. Spronck J.F.P., Szymkowiak A., 2013, PASP, 113, 1420
  \bibitem[Van Dokkum (2001)]{vanDokkum01}van Dokkum P.G., 2001, PASP 113, 1420
\bibitem[Vaytet, O'Brien \& Bode (2007)]{vaytet07}Vaytet N.M.H., O'Brien T.J., Bode M.F., 2007, ApJ, 665, 654
\bibitem[Vaytet et al. (2011)]{vaytet11}Vaytet N.M.H., O'Brien T.J., Page K.L., Bode M.F., Lloyd M., Beardmore A.P., 2011, ApJ, 740, 5
\bibitem[Verner et al. (1996)]{vern96}Verner D.A., Ferland G.J., Korista K.T., Yakovlev D.G., 1996, ApJ, 465, 487

\bibitem[Walter et al. (2012)]{fred12}Walter F.M., Battisti A., Towers S.E., Bond H.E., Stringfellow G.S., 2012, PASP, 124, 1057

\bibitem[Willingale et al. (2013)]{dick13}Willingale R., Starling R.L.C., Beardmore A.P., Tanvir N.R., O'Brien P.T., 2013, MNRAS, 431, 394
\bibitem[Wilms, Allen \& McCray (2000)]{wilms00}Wilms J., Allen A., McCray R., 2000, ApJ, 542, 914
\bibitem[Wolf et al. (2013)]{wolf13}Wolf W.M., Bildsten L., Brooks J., Paxton B., 2013, ApJ, 777, 136
\bibitem[Wolf, Townsend \& Bildsten (2018)]{wolf18}Wolf W.M., Townsend R.H.D., Bildsten L., 2018, ApJ, 855, 127
\bibitem[Woodward, Banerjee \& Evans (2020)]{wood20}Woodward C.E., Banerjee D.P.K., Evans A., 2020, Astron. Telegram, 13764
\bibitem[Woodward et al. (2019)]{wood19}Woodward C.E., Banerjee D.P.K., Evans A., Geballe T.R., Starrfield S., 2019, Astron. Telegram, 13096 
\bibitem[Woudt \& Ribeiro (2014)]{woudt14}Woudt P.A., Ribeiro V.A.R.M., eds, 2014, Astronomical Society of the Pacific Conference Series, Vol. 490, `Stella Novae: Past and Future Decades' (San Francisco: Astronomical Society of the Pacific)


\end{thebibliography}
\end{document}